\documentclass[fleqn,usenatbib]{mnras}

\usepackage{newtxtext,newtxmath}

\usepackage[T1]{fontenc}

\DeclareRobustCommand{\VAN}[3]{#2}
\let\VANthebibliography\thebibliography
\def\thebibliography{\DeclareRobustCommand{\VAN}[3]{##3}\VANthebibliography}

\usepackage{graphicx}	
\usepackage{amsmath}	

\usepackage{caption}

\usepackage{rotating}

\usepackage{float}

\title[Stellar wind of Cyg X-1]{The X-ray spectral-timing contribution of the stellar wind in the hard state of Cyg X-1}

\author[E. V. Lai \it{et al.}]{E. V. Lai,$^{1}$\thanks{E-mail: eleonorav@camk.edu.pl}
B. De Marco$^{1,2}$,
A. A. Zdziarski$^{1}$,
T. M. Belloni$^{3}$,
S. Mondal$^{1,3}$, P. Uttley$^{4}$,
\newauthor{V. Grinberg$^{5}$, J. Wilms$^{6}$, A. R\'o\.za\'nska$^{1}$}
\\
\\
$^{1}$Nicolaus Copernicus Astronomical Center, PAN, ul. Bartycka 18, Warsaw 00-716 Poland\\
$^{2}$Departament de Física, EEBE, Universitat Politècnica de Catalunya, Av. Eduard Maristany, 16, Barcelona  08019, Spain\\
$^{3}$INAF - Osservatorio Astronomico di Brera
Via E. Bianchi 46, I-23807 Merate, Italy \\
$^{4}$Anton Pannekoek Institute, University of Amsterdam, Science Park 904, 1098 XH Amsterdam, The Netherlands\\
$^{5}$European Space Agency (ESA), European Space Research and Technology Centre
(ESTEC), Keplerlaan 1, 2201 AZ Noordwijk, the Netherlands\\
$^{6}$Dr. Karl Remeis-Observatory, University of Erlangen-Nuremberg, Sternwartstr. 7, 96049 Bamberg, Germany
}

\date{Accepted XXX. Received YYY; in original form ZZZ}

\pubyear{2021}

\begin{document}
\label{firstpage}
\pagerange{\pageref{firstpage}--\pageref{lastpage}}
\maketitle

\begin{abstract}
The clumpy stellar wind from the companion star in high mass X-ray binaries causes variable, partial absorption of the emission from the X-ray source.
We studied XMM-Newton observations from the 7.22~d--long ``Cyg X-1 Hard state Observations of a Complete Binary Orbit in X-rays'' (CHOCBOX) monitoring campaign, in order to constrain the effects of the stellar wind on the short-timescale X-ray spectral-timing properties of the source.
We find these properties to change significantly in the presence of the wind. In particular, the longest sampled timescales (corresponding to temporal frequencies of $\nu\sim$~0.1--1~Hz) reveal an enhancement of the fractional variability power, while on the shortest sampled timescales ($\nu\sim$~1--10~Hz) the variability is suppressed. In addition, we observe a reduction (by up to a factor of $\sim$~1.8) of the otherwise high coherence between soft and hard band light curves, as well as of the amplitude of the hard X-ray lags intrinsic to the X-ray continuum. The observed increase of low frequency variability power can be explained in terms of variations of the wind column density as a consequence of motions of the intervening clumps. In this scenario (and assuming a terminal velocity of $v_{\infty}=2400\ {\rm km\ s^{-1}}$), we obtain an estimate of $l \sim$~0.5--1.5 $\times 10^{-4} R_{\ast}$ 
for the average radial size of a clump. On the other hand, we suggest the behaviour at high frequencies to be due to scattering in an optically thicker medium, possibly formed by collision of the stellar wind with the edge of the disc.
\end{abstract}

\begin{keywords}
X-rays: binaries; black hole physics; X-rays: individual (Cyg X-1); stars: wind, outflows; 
\end{keywords}



\section{Introduction}

Black hole X-ray binaries (BHXRBs) are observed to occasionally and recursively undergo dramatic changes of their X-ray spectral and timing properties \citep[e.g.][]{Homan_2005,Belloni_2010, Belloni_2011}. This behaviour is commonly ascribed to changes in the inner structure of the accretion flow \citep[e.g.][]{Gilfanov_2010}, and associated with hysteresis between two main accretion states, the so-called soft and hard states \citep[e.g.][]{Zdziarski_2004}. During the soft state most of the observed X-ray emission is consistent with being produced in a geometrically thin and optically thick disc \citep{Shakura_Sunyaev_1973}, and is characterized by a low level of rapid (from tens of seconds to milliseconds) X-ray variability (root mean square, rms, amplitude less than $\sim$~5 percent, e.g. \citealt{MunozDarias_2011}). In the hard state the X-ray spectrum is dominated by Comptonised emission. 
This is due to scattering of thermal photons from the optically thick disc \citep[e.g.][]{Shapiro_1976,Ichimaru_1977,Sunyaev_1979,Narayan_Yi_1994} and/or of synchrotron photons from the inner hot flow \citep[][]{Vurm_Poutanen2008,Veledina_2011} in a cloud of hot electrons, commonly referred to as hot corona, located close to the black hole (BH).
This state is characterized by high levels of rapid X-ray variability (rms even exceeding 30 percent, \citealt{MunozDarias_2011}) and a complex distribution over frequency of the X-ray variability power \citep[e.g.][]{Belloni_2005}.
While the phenomenology of these states is well established, the physical origin is still largely debated.

Cygnus X-1 (hereafter Cyg X-1) is amongst the best studied X-ray binary systems. It is composed of a BH with $M_{\odot}=21.2 \pm 2.2\,\rm{M_{\odot}}$ and its supergiant $\rm{O}9.7\,\rm{Iab}$ companion star (HDE 226868) with $M_{\ast}=40.6^{+7.7}_{-7.1}\,\rm{M_{\odot}}$ and $R_{\ast}=22.3\pm1.8\,\rm{R_{\odot}}$ \citep{MillerJones_2021}. The system is characterised by a quasi-circular orbit with an orbital period of $5.599829(16)\,\rm{d}$ \citep{Gies_2003}. 
Due to the size of the companion, this system is classified as a high mass X-ray binary, where the BH accretes mass via the strong stellar wind of the supergiant. 

Like all other BHXRBs, Cyg X-1 shows complex temporal frequency dependence of its X-ray variability power, with a power spectral density function (PSD) characterized by two main broad ``humps'' in the hard state, 
that change drastically to a smoother profile in the soft state \citep[e.g.][]{Nowak_1999,Pottschimdt_2000,Pottschmidt_2003,Axelsson_2005a,Boeck_2011a,Grinberg_2014}. In addition, highly coherent hard ($E\gtrsim1$~keV) and soft ($E\lesssim1$~keV) X-ray band variations are regularly 
observed, whereby variable hard photons lag behind soft photons with a delay of order 1 percent of the variability timescale \citep[hereafter ``hard lags'', e.g.][]{Nowak_1999,Grinberg_2014}. Various mechanisms have been proposed to explain the observed X-ray spectral-timing behaviour of BHXRBs \citep{Poutanen_1999,Misra_2000}, with models of inward propagation of mass accretion rate fluctuations currently offering the most compelling physical explanation \citep{Lyubarskii_1997,Kotov_2001,Arevalo_Uttley2006,Ingram_vanderKlis2013,Mushtukov_2018,Mahmoud_2018_A,Mahmoud_2018_B, Bollimpalli_2020}.

However, an important additional component of high mass X-ray binary systems like Cyg X-1 is the stellar wind from the companion, early type, supergiant star.
Evidence for the presence of a stellar wind was soon found in Cyg X-1 \citep[e.g.,][]{Li_1974,Remillard_1984,Balucinska_2000,Miller_2005a,Poutanen_2008,Hanke_2009a,Grinberg_2015,Miskovicova_2016}.
The strong line-driven wind presents an inhomogeneous structure, with clumps characterized by highly dense regions \citep{Owocki_1984,Owocki_1988,Feldmeier_1995}. When a clump intercepts our line of sight to the X-ray source, the X-ray emission is partially absorbed by the wind material leading to a dipping event (sudden drop of the X-ray flux). Such events occur more frequently and are more intense at superior conjunction, i.e. when the compact object lies behind the donor star (at orbital phase $\phi_{\mathrm{orb}}=0$ and our line of sight probes regions of the stellar wind closer to the surface of the companion, e.g. \citealt[][]{Balucinska_2000,Grinberg_2015,Miskovicova_2016}).

It is reasonable to expect that wind absorption does not only affect the observed spectrum, but also the observed X-ray variability properties of the source \citep{Grinberg_2015,Grinberg_2020}. 
For example, motions of wind clumps through our line of sight to the X-ray source are expected to produce X-ray spectral-timing signatures on the relevant timescales \citep{ElMellah_2020}. Given current estimates of the size of the clumps, such timescales can be of the order of minutes or shorter \citep{Grinberg_2017,Hirsch_2019}. In addition, the wind is found to be affected by the radiation field around the BH, resulting in stronger absorption during the hard state than during the soft state \citep{Boroson_Vrtilek_2010,Nowak_2011}. Given the strong flux variability characterising the hard state of BHXRBs \citep{MunozDarias_2011,Heil_2012}, additional wind variability due to changes in the irradiating flux might be expected \citep{Nicastro_1999,Krongold_2007,Silva_2016}. 

All this suggests that the observed X-ray spectral-timing characteristics of Cyg X-1 can be significantly influenced by the stellar wind, thus affecting any study aimed at constraining the intrinsic properties of the X-ray source. On the other hand, the X-ray spectral-timing signatures characteristics of the most wind-absorbed phases can be a powerful tool to determine the physical properties of the wind \citep[e.g.][]{Grinberg_2015,Silva_2016,DeMarco_2020,ElMellah_2020}.

Cyg X-1 has been recently the target of an unprecedented long multiwavelength monitoring campaign, the ``Cyg X-1 Hard state Observations of a Complete Binary Orbit in X-rays'' (CHOCBOX) campaign, with XMM-Newton as the primary instrument. The XMM-Newton monitoring consists of four observations of the source in its hard state, for a total exposure of about $\sim$~572~ks  
and covering about one and a half orbital periods (7.22~d). In this paper, we study this long dataset, focusing on determining how the variability of the stellar wind influences the X-ray spectral-timing properties of the source.

\section{Data reduction}{\label{data}}

For the spectral-timing measurements, which are optimised for high count rates, we consider only the EPIC-pn data. The four observations analysed were carried out between 2016 May 27 and June 2, with a total exposure of $\sim$~572~ks  (before data screening). The analysed data are acquired in Timing mode.
The log of the observations is reported in Tab.\ref{tab:table}. Hereafter, we will refer to each observation using the last three digits of their ObsID (as specified in Tab.\ref{tab:table}). 

The data reduction has been carried out using the XMM-Newton Science Analysis System (SAS) software (version 16.9.0), following standard procedures. 
In order to identify and filter out time intervals affected by high particle background, we extracted light curves with a time resolution of 1~s in the energy range 10--15~keV (where the instrument response drops). We checked the light curves to identify background flares, but none was found. 

The analysed observations are affected by telemetry drop-outs, i.e. periods of high count rate producing buffer overflow, during which photons are not collected. These periods result in gaps in the data, which are accounted for in the selection of Good Time Intervals, GTIs, using the SAS task \texttt{tabgtigen}. The resulting GTIs are very short during all observations, with an average length of $\sim$~10~s.

We extracted source counts in the range RAWX:30--46. Using the SAS task \texttt{epatplot}, we checked for the presence of pile-up. We found that the data are affected by some fraction of pile-up, thus we mitigated it by excluding the central (RAWX:36--39) pixels. 
Since the observations are in Timing mode, it is not possible to select a source-free region for extraction of background events, while subtracting the background from the outer columns (RAWX intercepting the tails of the PSF) may modify the source spectrum \citep[]{Ng_2010}. However, the source is very bright (with a 0.5--10~keV average count rate of $\sim$~220~counts/s), thus the effects of background are expected to be negligible \citep[e.g.][]{Ng_2010}. Finally, we notice that the most absorbed periods, thus potentially more affected by the background, are excluded in our spectral analysis (Sect.~\ref{CS}), while the background is not expected to contribute to the rapid X-ray variability of BHXRBs \citep[e.g.][]{Uttley_2011,DeMarco_2017}. Therefore, we decided not to subtract the background.

Ancillary response files (ARF) and redistribution matrix files (RMF) were extracted using the SAS tools \texttt{arfgen} and \texttt{rmfgen}. We generated an ARF for the full region (RAWX:30--46) and a second ARF for the excluded region (RAWX:36--39). The ARF of our extraction region was obtained by subtracting the latter from the former using the command \texttt{addarf} \citep[e.g.][]{Wilkinson_Uttley2009}.
We used calibration files (CCF) as of June 2019.
The spectra were re-binned to have a minimum of 20 counts in each energy bin.
For the spectral fits (Sect.~\ref{CS}) we used Xspec v12.10.1 \citep{Arnaud_1996}, while codes for the spectral-timing analysis (Sects.~\ref{PS} and \ref{CS}) were implemented using Python 3.6.

\begin{table}
	\centering
    \caption{Log of the XMM-Newton EPIC-pn observations analysed. Effective exposures after regular data screening (Total exposure) and after the additional selection of time intervals not affected by wind absorption (NWA exposure) are reported. The orbital phase $\phi_{\mathrm{orb}}$ is computed using the ephemeris reported in \citet{Gies_2008}. The last three digits of the ObsID used throughout the test to identify each observation are marked in boldface.}
	\label{tab:table}
	\begin{tabular}{ccccc} 
		\hline
		ObsID & Date & Total & NWA & $\phi_{\mathrm{orb}}$ \\
		      &      & exposure  & exposure & \\
		& (yyyy--mm--dd) & (ks) & (ks) & \\
		\hline
		0745250\textbf{201} & 2016--05--27 & 92.9 & 30.9 & 0.82--0.06\\
		0745250\textbf{501} & 2016--05--29 & 83.1 & 79.8 & 0.17--0.46\\
		0745250\textbf{601} & 2016--05--31 & 76.7 & 74.8 & 0.53--0.79\\
		0745250\textbf{701} & 2016--06--02 & 84.6 & 44.9 & 0.89--0.11\\
		\hline
	\end{tabular} 
\end{table}
\begin{figure*}
\includegraphics[trim={0cm 0cm 0cm 0cm},width=1\textwidth]{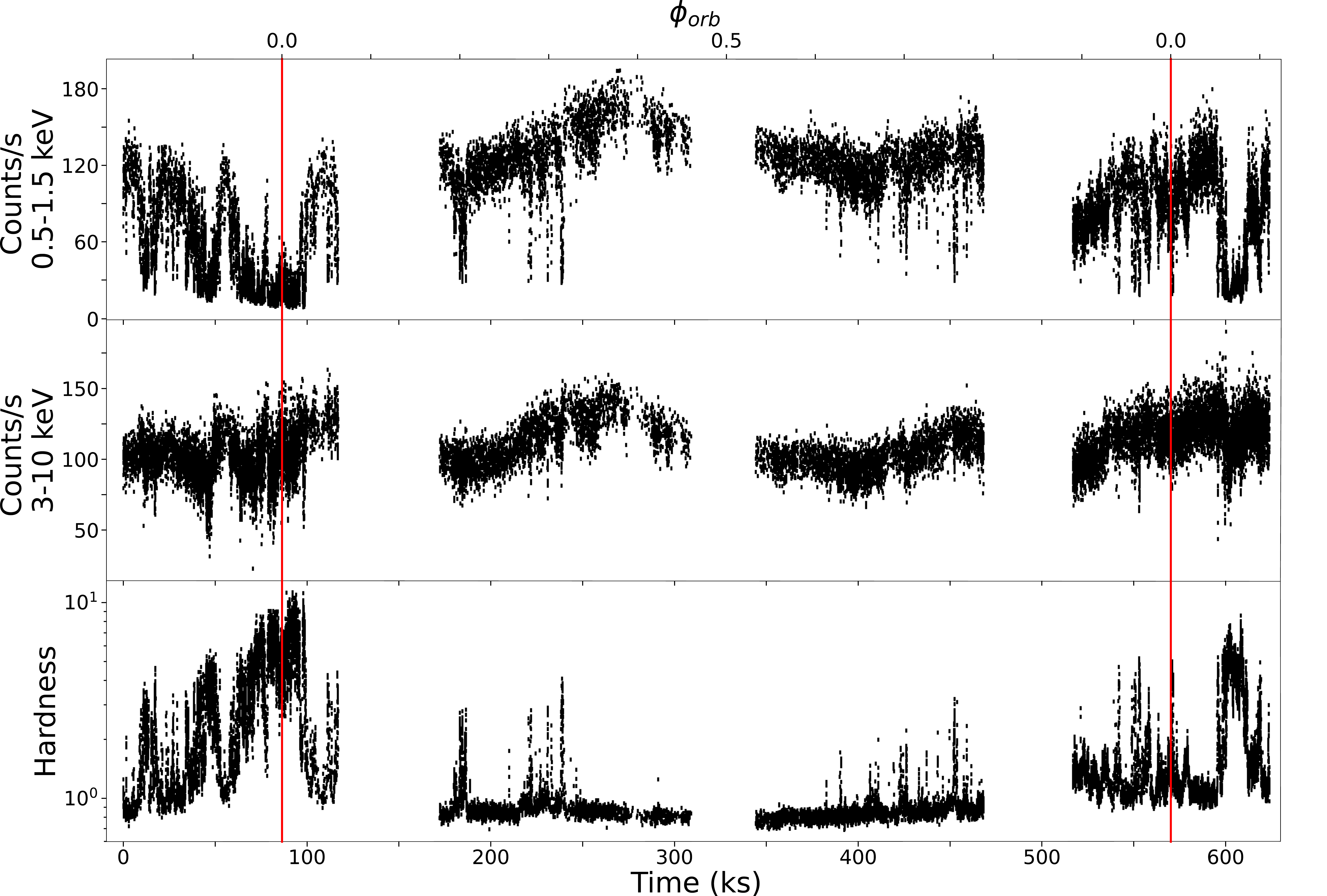}
\caption{XMM-Newton EPIC-pn light curves of the four analysed observations of Cyg X-1 (i.e. observations 201, 501, 601, and 701, each separated by the gaps in the light curve). The reference starting time of the monitoring is 57535.0 MJD. The upper panel reports the 0.5--1.5~keV light curve. The middle panel reports the 3--10~keV light curve. The bottom panel reports the ratio between count rates in the 3--10~keV and 0.5--1.5~keV energy bands. The red lines indicate the two consecutive passages at superior conjunctions ($\phi_{\mathrm{orb}} = 0 $)  occurring during the XMM-Newton monitoring.}
\label{HIDiagram}
\end{figure*}

\subsection{Selection of events not affected by wind absorption}\label{selezione_NWA}
 
Depending on the orbital phase of the compact object the X-ray emission will be more or less absorbed by the stellar wind. As a consequence of the orbital modulation of the intervening absorbing column, light curves show recurrent dips of variable intensity. Due to the spectral dependence of the opacity of the wind, the soft X-ray band is the most affected.

Fig.~\ref{HIDiagram} shows the EPIC pn light curves of the source extracted (with a time bin of 10~s) in the two energy bands: 0.5--1.5~keV and 3--10~keV. During the monitoring, two consecutive passages at superior conjunction (occurring during observations 201 and 701) are observed. These are marked by red vertical lines in Fig.~\ref{HIDiagram}. 
The 0.5--1.5~keV light curve  (Fig.~\ref{HIDiagram}, top panel) shows several dips, in particular around superior conjunction. Such dips are less intense in the 3--10~keV energy band (Fig.~\ref{HIDiagram}, middle panel), resulting in a net increase of the spectral hardness
(defined as the ratio between count rates in the 3--10~keV and 0.5--1.5~keV energy bands; Fig.~\ref{HIDiagram}, bottom panel).
We note that, though clustering around superior conjunction (at the beginning and at the end of the monitoring), dips are sparsely present throughout the entire campaign.

\begin{figure*}
\includegraphics[trim={8cm 0 9cm 0},width=1.0
\textwidth]{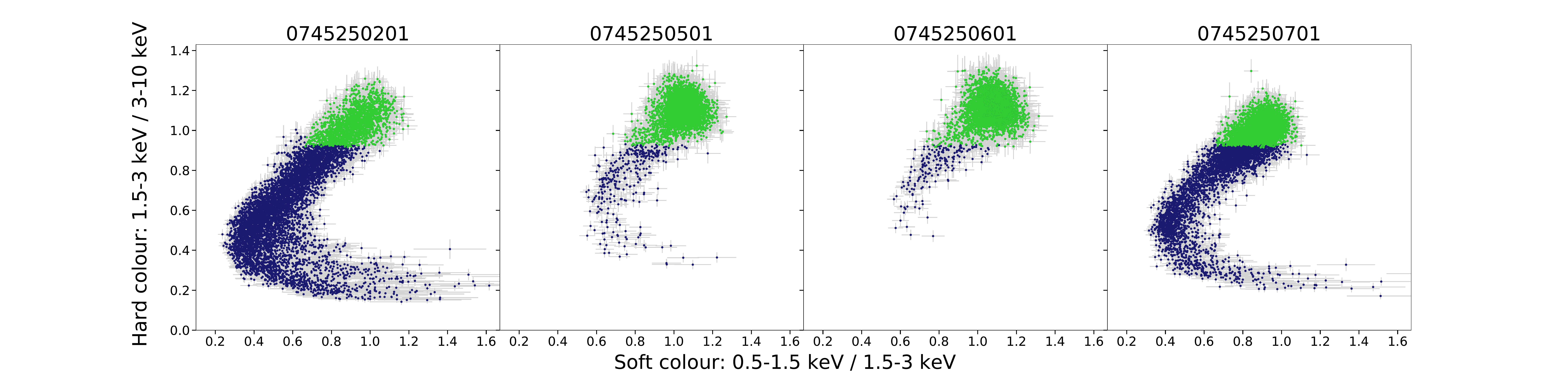}
\caption{The colour-colour diagram of each XMM-Newton observation, showing hard and soft colours calculated for soft (0.5--1.5~keV), intermediate (1.5--3~keV) and hard bands (3--10~keV). Each point in the plot corresponds to a 10~s--long data segment. 
Time intervals in green are those characterised by values of hard colour $\geqslant$~0.95 and soft colour $\geqslant$~0.7, thus representing our selection of NWA datasets, the least affected by wind absorption.}
\label{fig:CCDiagram}
\end{figure*}

We followed the method proposed in \cite{Nowak_2011} and \cite{Hirsch_2019} to select time intervals not affected by absorption. 
The method consists in constructing a colour-colour diagram which shows the time-resolved spectral behaviour of the source. The diagram displays ratios between a soft and an intermediate band (soft colour) and between an intermediate and a hard band (hard colour).  
The position of the source in the colour-colour diagram depends on the amount of absorption along the line of sight \citep{Grinberg_2020}.

In order to build the colour-colour diagram, we used light curves with a time bin of  10~s so as to uncover also very short dips.
We considered the three energy bands 0.5--1.5~keV, 1.5--3~keV, and 3--10~keV (hereafter soft, intermediate, and hard). The resulting colour-colour diagram for each XMM-Newton observation is shown in Fig.~\ref{fig:CCDiagram}.

Following \cite{Hirsch_2019}, time intervals characterised by high values of both hard and soft colours (upper right corner of the colour-colour diagram) are the least affected by wind absorption.
In order to extract a dataset relatively free from wind absorption, we thus selected events from that region of the diagram by defining a threshold for the hard and soft colours.  
We selected all the 10~s--long data segments with a hard colour $\geqslant$~0.95 and a soft colour $\geqslant$~0.7. 
We consider the simple model adopted in \cite{Hirsch_2019}, which comprises a power law with spectral index $1.7$ partially covered by a neutral absorber (representing the clumpy wind), modified by Galactic absorption (\texttt{tbnew $\times$ tbpcf $\times$ powerlaw} in Xspec, with Galactic $\mathrm{N_H} = 0.7\times 10^{22}\,\mathrm{cm^{-2}}$; \citealt{Basak_2017,HI4PIcoll}; and assuming abundances from \citealt{Wilms_2000}\footnote{https://pulsar.sternwarte.uni-erlangen.de/wilms/research/tbabs/}). For this model, we find that if the partial covering factor is kept constant and only the wind column density varies, the theoretical curve that best samples the range of hard and soft colours covered by the data corresponds to a covering factor of $\sim$~0.9. For such a value of the covering factor, the chosen threshold of hard and soft colours selects data characterised by a wind column density $\mathrm{N_{H, w}}$ $\lesssim1.08\times 10^{22}\,\mathrm{cm^{-2}}$. The final datasets resulting from the described filtering process will be hereafter referred to as ``NWA'' (standing for ``no wind absorption'') as opposed to ``Total'', which will be used to refer to the datasets unfiltered from wind absorption.
It is worth noting that the inferred theoretical curve describing the colour-colour tracks and the corresponding physical parameters represent just an approximation. Indeed, the curve does not accurately reproduce the observed tracks (Appendix \ref{APP_CCDiagram_model_data}). 
As discussed in \cite{Grinberg_2020}, this highlights the need for more complex absorption models in order to properly describe these tracks. While this is beyond the scope of this paper, this issue will be addressed in a follow up paper. 

It is worth noting that our filtering choices, together with the use of a 10~s time bin for the light curves employed to build the colour-colour diagram, ensure a good filtering of strong wind absorption events of duration $\geqslant$~10~s. Nonetheless, residual absorption might still be present. Wind clumps characterised by lower $\mathrm{N_{H, w}}$ and/or lower covering factor, and moving faster across the line of sight (thus producing shorter dips) would be the major contributors to this residual absorption.


\section{Power Spectra}{\label{PS}}

In order to study the effects of the stellar wind on the X-ray variability of the source we first extracted the PSD of the NWA and Total datasets of each observation. The PSD were extracted in the soft, intermediate and hard energy bands defined in Sect.~\ref{selezione_NWA}. 
To this aim we used light curves with a time resolution of 6~ms. The light curves were split into segments, containing 1666 data points each (which correspond to a length of 9.996~s).
This allows us to sample the range of temporal frequencies $\nu\sim$~0.1--83~Hz (corresponding to variability timescales between $\sim$~0.01--10~s). The chosen segment length is the result of a trade-off between obtaining a sufficiently broad frequency coverage while retaining a high number of GTIs.
We calculated the PSD of each segment and then averaged over all segments. We adopted the fractional rms normalization \citep{Belloni_1990, Miyamoto_1991}. We did so for each observation separately, in order to study the evolution of the wind-modulated variability as a function of the orbital phase. 
We estimated the Poisson noise contribution by fitting the PSD with a constant at frequencies $>25$~Hz. We corrected the PSD for Poisson noise by subtracting this contribution. The PSD was geometrically rebinned, using a rebinning factor of $\nu_{i+1}= 1.2\,\nu_{i}$.

\begin{figure*}
\centering
\includegraphics[trim={4cm 0 8cm 0.3cm},width=1.0\textwidth]{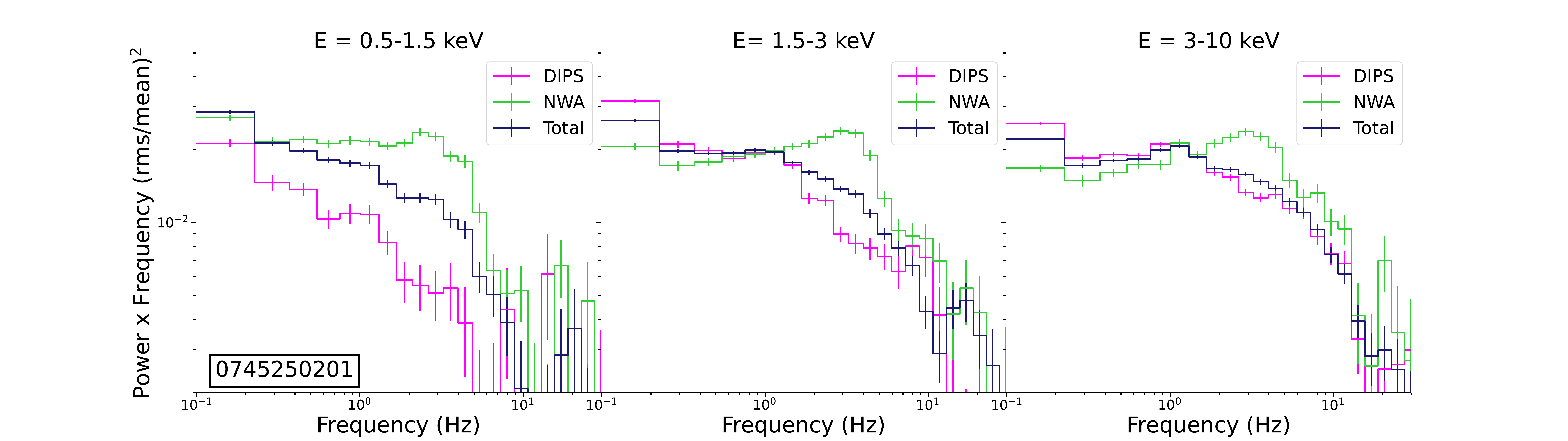}
\includegraphics[trim={4cm 0 8cm 0.3cm},width=1.0\textwidth]{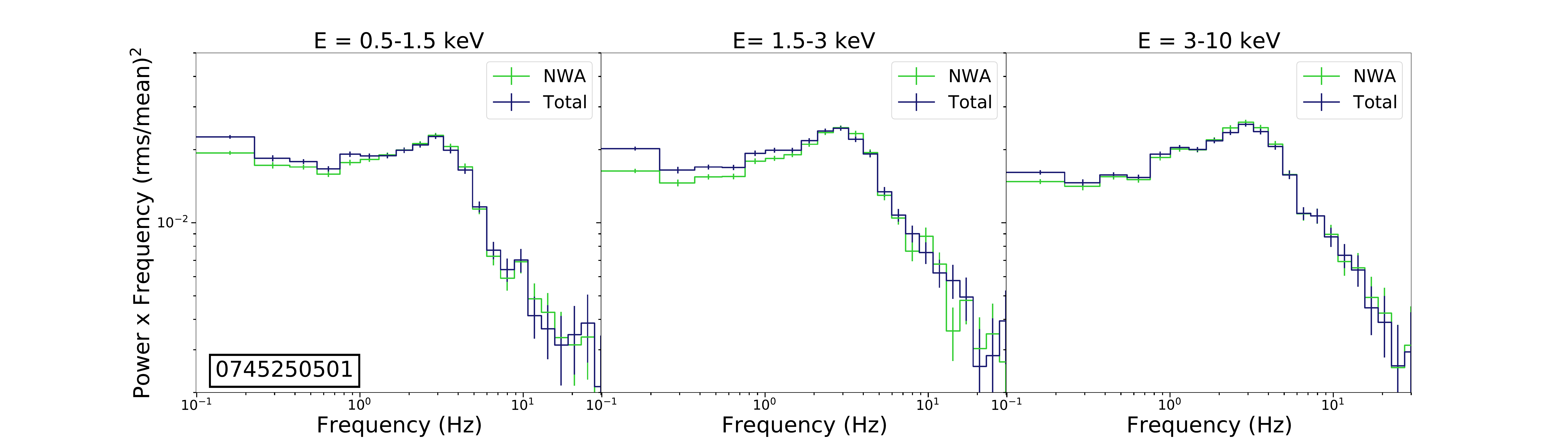}
\includegraphics[trim={4cm 0 8cm 0.3cm},width=1.0\textwidth]{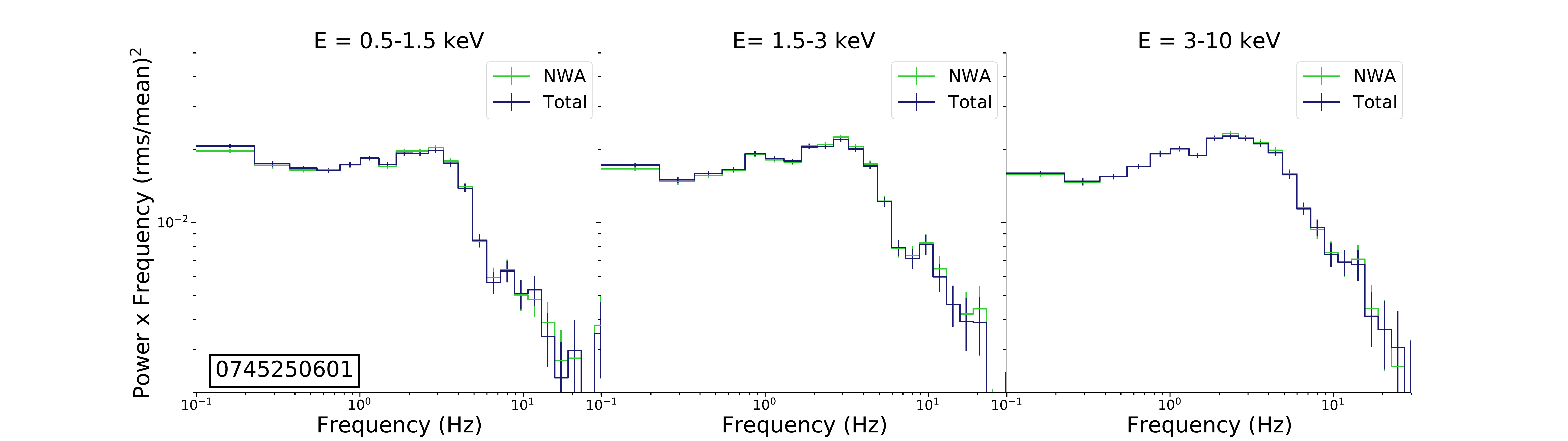}
\includegraphics[trim={4cm 0 8cm 0.3cm},width=1.0\textwidth]{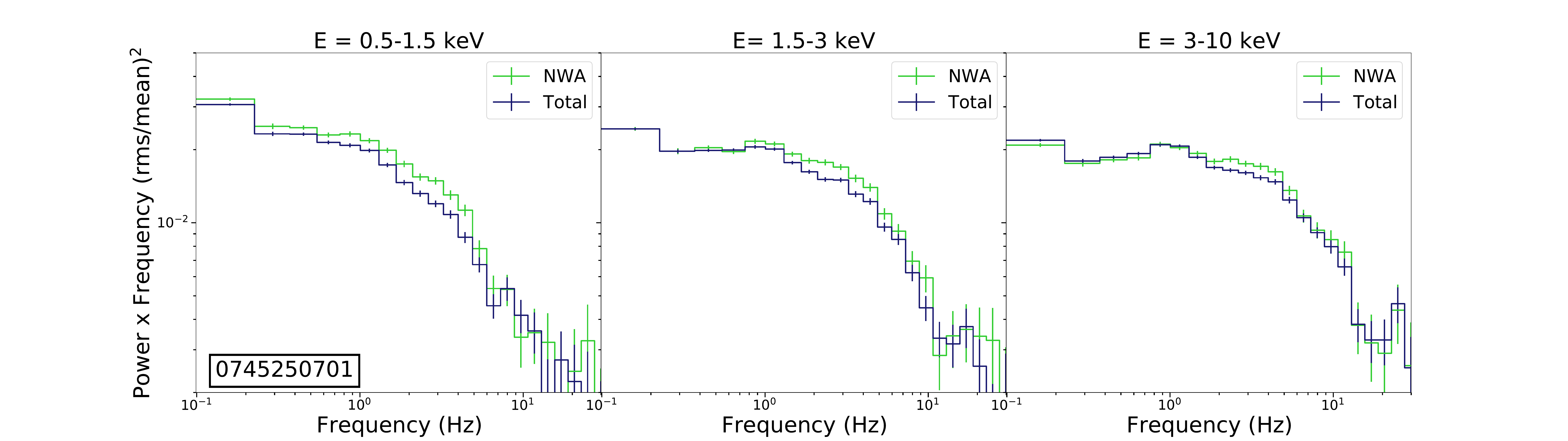}
\caption{PSD for the Total (blue), the NWA (green) datasets of each observation (from top to bottom) in the soft (0.5--1.5~keV), intermediate (1.5--3~keV) and hard (3--10~keV) energy bands (from left to right). For observation 201 we also show the PSD obtained from selection of the wind-dominated (characterized by a hard colour $\leqslant$~0.8, Sect.~\ref{DIPS_section}) ``DIPS'' dataset (in magenta).} 
\label{psd}
\end{figure*}

In Fig.~\ref{psd}, we compare the PSDs obtained from the NWA and Total datasets. The PSDs are shown separately for each observation and for the soft, intermediate, and hard energy bands. 
We observe that the wind contributes significantly to the short-timescale X-ray variability. The main effect of the wind is that of smoothing out the typical double-hump shape of the PSD of Cyg X-1 in the hard state \citep[e.g.][]{Pottschmidt_2003,Axelsson_Done_2018,Mahmoud_2018_A,Mahmoud_2018_B}. This shape is recovered when considering the NWA datasets and when the source is not at superior conjuction (thus it is less affected by wind absorption). 
More specifically, wind absorption tends to reduce (by up to a factor of $\sim$~2) the fractional variability at high frequencies ($\gtrsim$~1~Hz), and to increase it at low frequencies ($\lesssim$~1~Hz), as most clearly seen in observation 201.
However, given the limited bandpass, the effects of the wind at low frequencies are more difficult to constrain.

Notably, the PSD of observation 701 remains quite smooth also after filtering out wind-absorbed segments. We verified that this is due to residual wind absorption. Indeed, when choosing a tighter threshold of hard and soft colours for the selection of the NWA dataset (Sect.~\ref{selezione_NWA}), the double-hump shape appears (Appendix \ref{APP_701}). This different selection also allows for additional low-frequency variability due to the wind in the Total dataset to be clearly observed in this observation as well.
The need for a different threshold suggests that the average properties of the wind may be different between consecutive passages at superior conjunction (as also indicated by the different extension of the colour-colour track towards low values of soft colours and the spread around the main track\footnote{We note that in general, spectral hardening and/or softening of the source may also produce such changes in the colour-colour tracks (e.g. see figure 2 of \citealt{Grinberg_2020}). However, our best-fit models do not show significant difference in spectral slope between these two observations, Table~\ref{tab_spectra}.}, during observations 201 and 701, Fig.~\ref{fig:CCDiagram}).
Nevertheless, a tighter threshold for the selection of NWA events in observation 701 severely reduces the amount of usable exposure (thus decreasing the signal-to-noise of spectral-timing measurements). For this reason, and for consistency with the other observations, we decided to use the same selection criteria as for the other observations (Sect.~\ref{selezione_NWA}).

Finally, it is worth noting that the effects of the wind on the PSD of the source are not limited to the soft (and more absorbed) energy bands, but can be observed up to the highest sampled energies (i.e. $E=3$--10~keV), suggesting rather high column densities.

\section{Cross spectral analysis}\label{CS}

Cross-spectral analysis informs us about the amount of correlated variability and the causal relationship between different spectral components \citep[e.g.][]{Nowak_1999,Uttley_2014}. It is well-known that in the hard state of BHXRBs the primary Comptonisation component and the disc vary in a linearly correlated way, showing high levels of coherence on a broad range of timescales \citep[e.g.][]{Wilkinson_Uttley2009}. In addition, the disc is observed to lead the variations of the Comptonisation component on the long timescales ($>1$~s, \citealt{Uttley_2011}), while, for some sources, it is observed to respond to hard X-ray variability on the shortest sampled timescales ($<1$~s) \citep{Uttley_2011,DeMarco_2015,Kara_2019,DeMarco_2021}. The former behaviour is usually interpreted in terms of inward propagation of mass accretion rate fluctuations in a spectrally inhomogeneous medium \citep{Lyubarskii_1997,Kotov_2001,Arevalo_Uttley2006,Ingram_vanderKlis2013,Mushtukov_2018,Bollimpalli_2020}, while the latter is ascribed to the thermal response of the disc to variable hard X-ray irradiation (thermal reverberation, e.g. \citealt{Uttley_2011,DeMarco_2015,DeMarco_2016,DeMarco_2017,Kara_2019,Wang_2020,DeMarco_2021}).  
In this section, we analyse the X-ray cross-spectral timing properties of Cyg X-1 in order to study the correlations and causal relationship between the disc and the primary Comptonisation components, and estimate the effects of the wind on such properties.

We first identified the energy bands where the disc and the Comptonisation component dominate by fitting the time-averaged spectra of the source. In order to minimise the effects of the wind on the primary X-ray continuum, the spectra were extracted considering only the NWA dataset. The time-averaged spectra of all observations were fit jointly.
We discarded data below 0.7~keV in order to avoid distortions due to electronic noise in Timing mode (calibration file: XMM-CCF-REL-265, Guainazzi, Haberl and Saxton 2010)\footnote{https://www.cosmos.esa.int/web/xmm-newton/ccf-release-notes}. 

The model used for the fit is \texttt{TBnew $\times$  [diskbb + nthComp + relxillCp]} in Xspec. It includes a multicolour disc component, \texttt{diskbb},  \citep{Mitsuda_1984},  and the soft excess, well modelled by a soft Comptonisation component, \texttt{nthComp} \citep{Zdziarski_1996,Zycki_1999}. Even though difficult to physically interpret, this component is necessary to obtain a reasonable fit \citep[e.g.][]{Zdziarski_2004,Basak_2017}.
The \texttt{relxillCp} \citep{Dauser_2014,Garcia_2015} component has been used to model the hard Comptonisation component and its associated reflection spectrum from the inner disc. 
Among the different spectra, we tied the inclination and the column density of the ISM to be the same for each observation, whereas for every single spectrum, we tied the seed photon temperature to the inner disc temperature and we fixed the high-energy cut-off of the hard Comptonisation component at 100~keV \citep{Basak_2017}.
The column density of the ISM was left free to vary in order to account for possible residual wind absorption. Indeed, we found $\mathrm{N_H} \sim 0.9\,\times 10^{22}\,\mathrm{cm^{-2}}$, slightly higher than the value measured in Galactic neutral atomic Hydrogen column density surveys (i.e. $0.7\,\times 10^{22}\,\mathrm{cm^{-2}}$, \citealt{HI4PIcoll}).
We observe significant residuals at energies between 1.5 and 2.5~keV, i.e. around the absorption edges of the detector, likely associated with incorrect calibration. Indeed, after ignoring this energy range the fit significantly improves, with a $\chi^2 / dof = 11582/8268$.  
We found an inner disc temperature in the range $kT_{in} \sim$~0.17--0.2~keV and a spectral index for the hard Comptonisation component $\Gamma_H \sim$~1.3--1.4 (see Table \ref{tab_spectra}). 

\begin{figure}
\includegraphics[trim={0cm 0cm 0cm 0cm},width=0.47\textwidth]{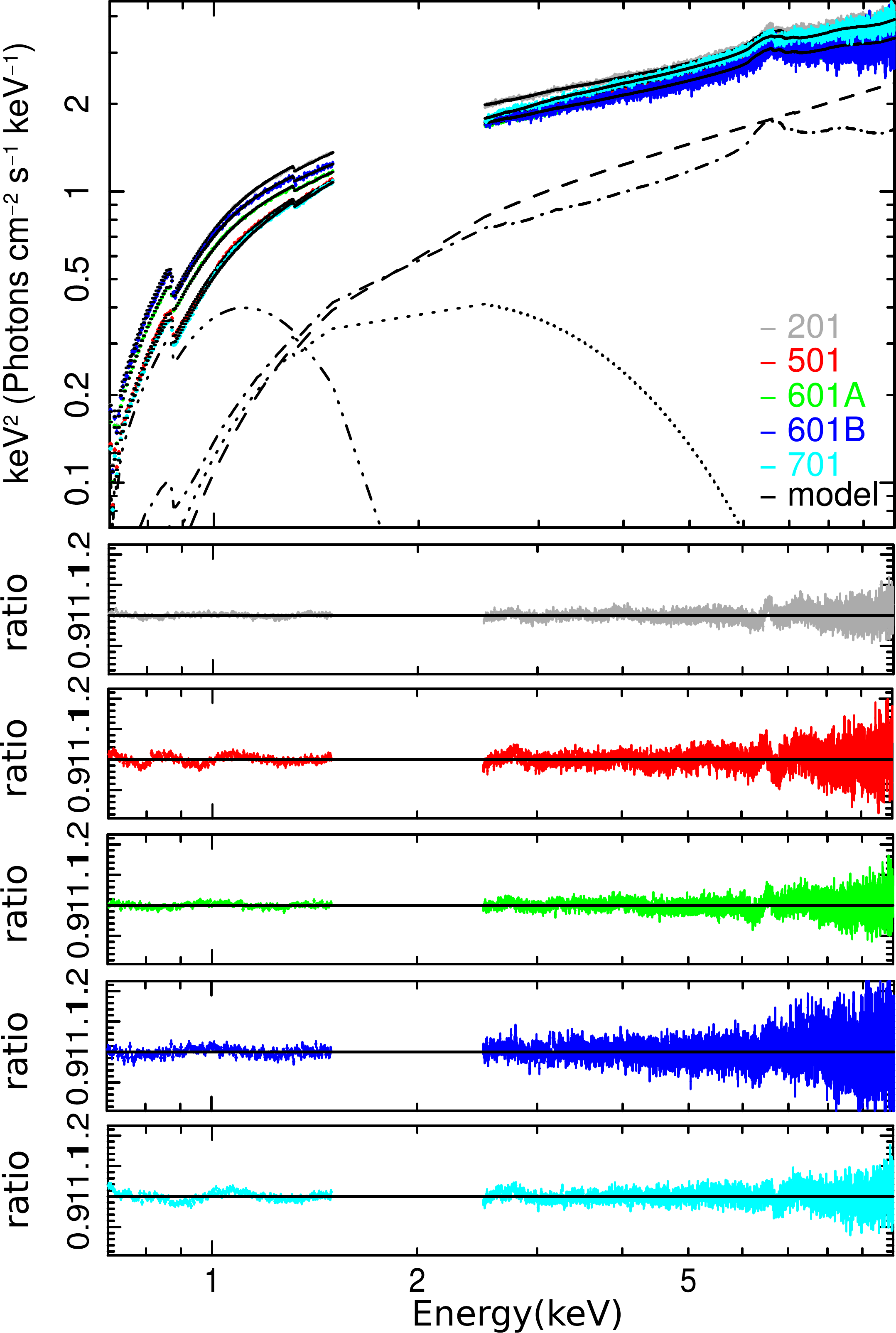}
\caption{The best-fit models from the joint fit of the NWA time-averaged spectra of all the observations (observation 601 is divided in two due to a gap in the data). The best-fit model (\texttt{Tbnew $\times$ [diskbb + nthComp + relxillCp]}) is overplotted in black (solid curve) for each observation. For clarity, the single components of the model are reported only for observation 201: triple-dot-dashes correspond to the disc blackbody (\texttt{diskbb}), dots to the soft excess (\texttt{nthComp}), dashes to the hard Comptonisation component (\texttt{relxillCp} Comptonised part only) and dot-dashes to its reflection with the disc (\texttt{relxillCp} reflection part only). There are large instrumental residuals in the energy range 1.5--2.5~keV, so this range is excluded from the fit. The bottom panels show the ratios of the data to the best-fit model for each spectrum separately.}
\label{spettro}
\end{figure}

In Fig.~\ref{spettro} we show the spectrum, the best-fit model and their ratio for each observation.
We see that the disc dominates below $\sim$~1~keV and that the 2--10~keV energy band is dominated by the hard X-ray primary emission. 
Thus we decided to use the energy bands 0.3--1~keV and 2--10~keV for the computation of cross spectra.
We extracted light curves in these two energy bands, using the same sampling parameters (time bin and segments length) as used for the computation of the PSDs (Sect.~\ref{PS}).
For each light curve segment, we computed the cross-spectrum, then we averaged over the different cross spectra. 
Following \cite{Ingram_2019}, we rebinned the cross spectra in order to have a minimum number of 500 points per frequency bin to obtain good S/N, so to avoid the complexities of defining the Poisson noise contribution.
This procedure was repeated separately for the NWA and Total datasets of each observation.
We used the average cross spectra to compute time lags and coherence as detailed in Sects.~\ref{cohe_freq} and \ref{lag_freq}.

\subsection{Coherence as a function of frequency}
\begin{figure*}
\centering
\hspace*{-0.3cm}
\includegraphics[width=0.5\linewidth]{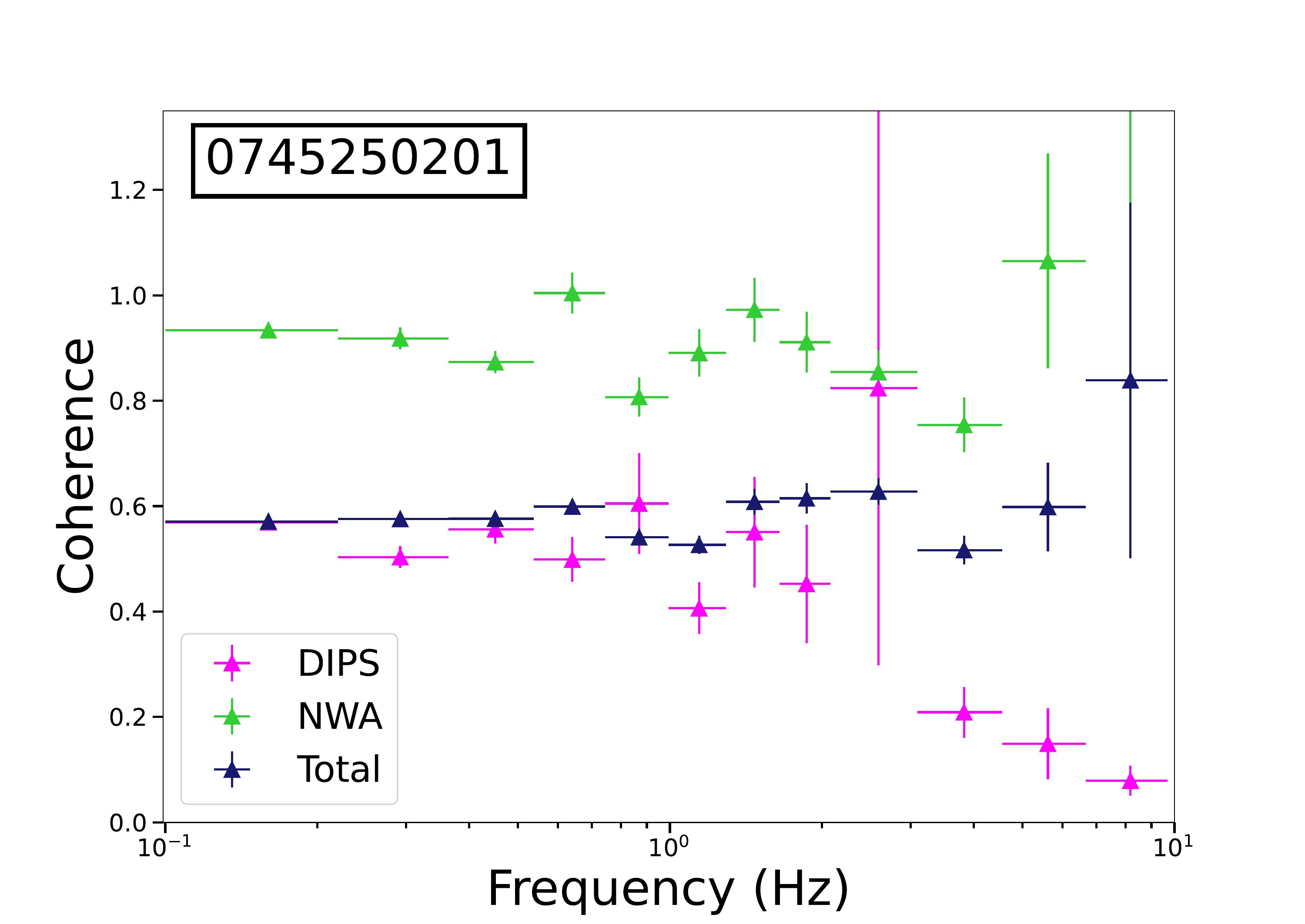}
\includegraphics[width=0.5\linewidth]{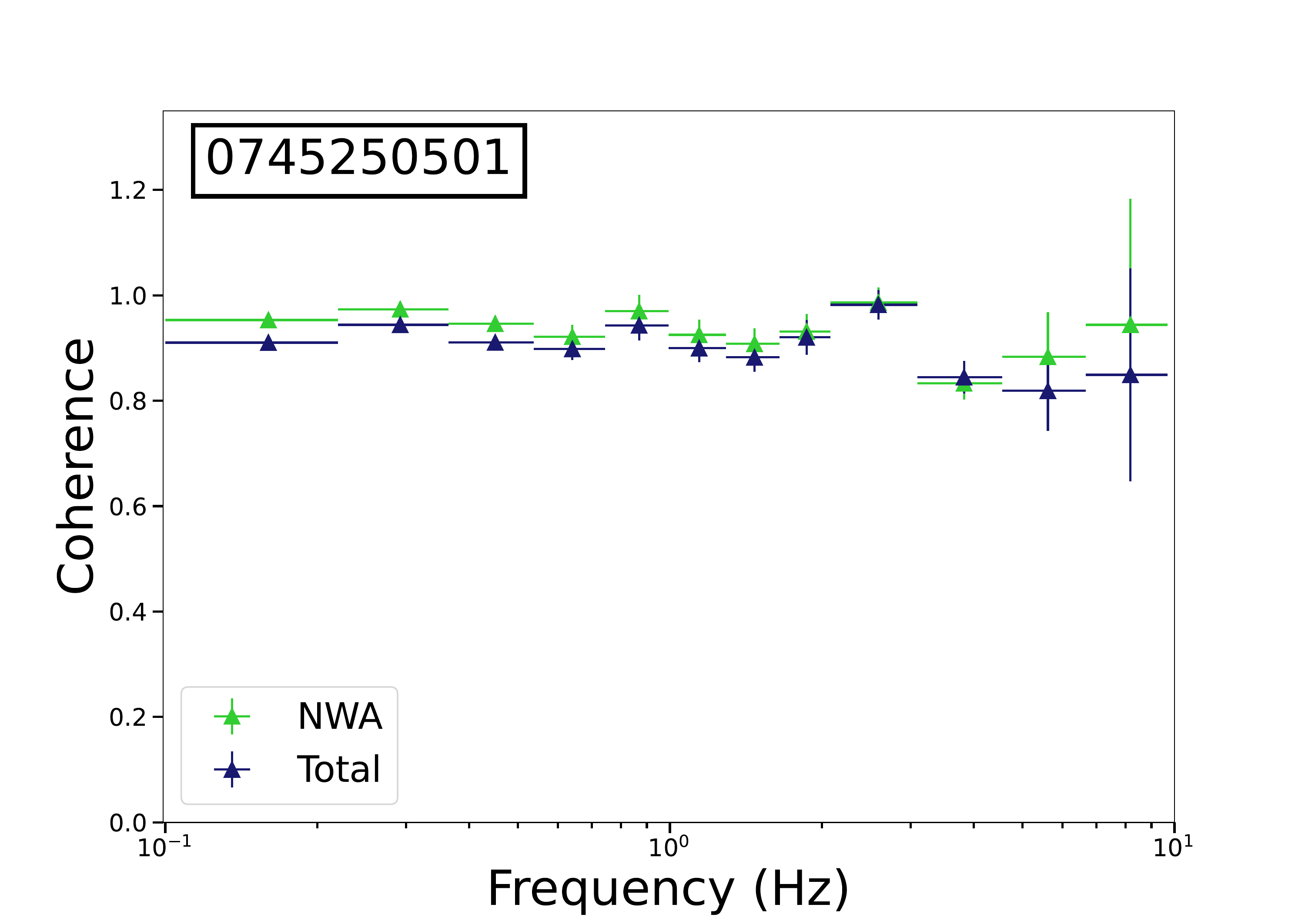}
\hspace*{-0.3cm}
\includegraphics[width=0.5\linewidth]{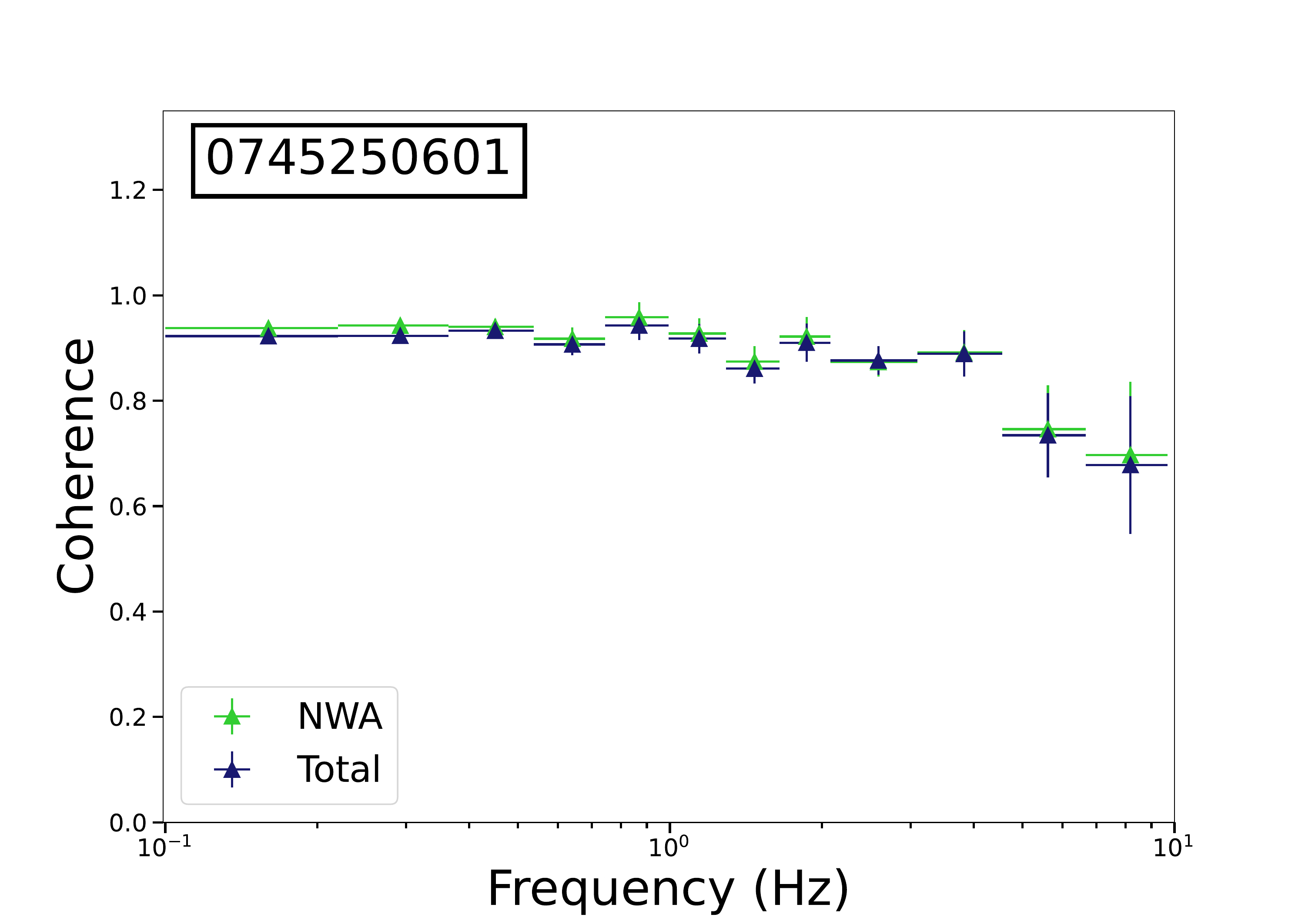}
\includegraphics[width=0.5\linewidth]{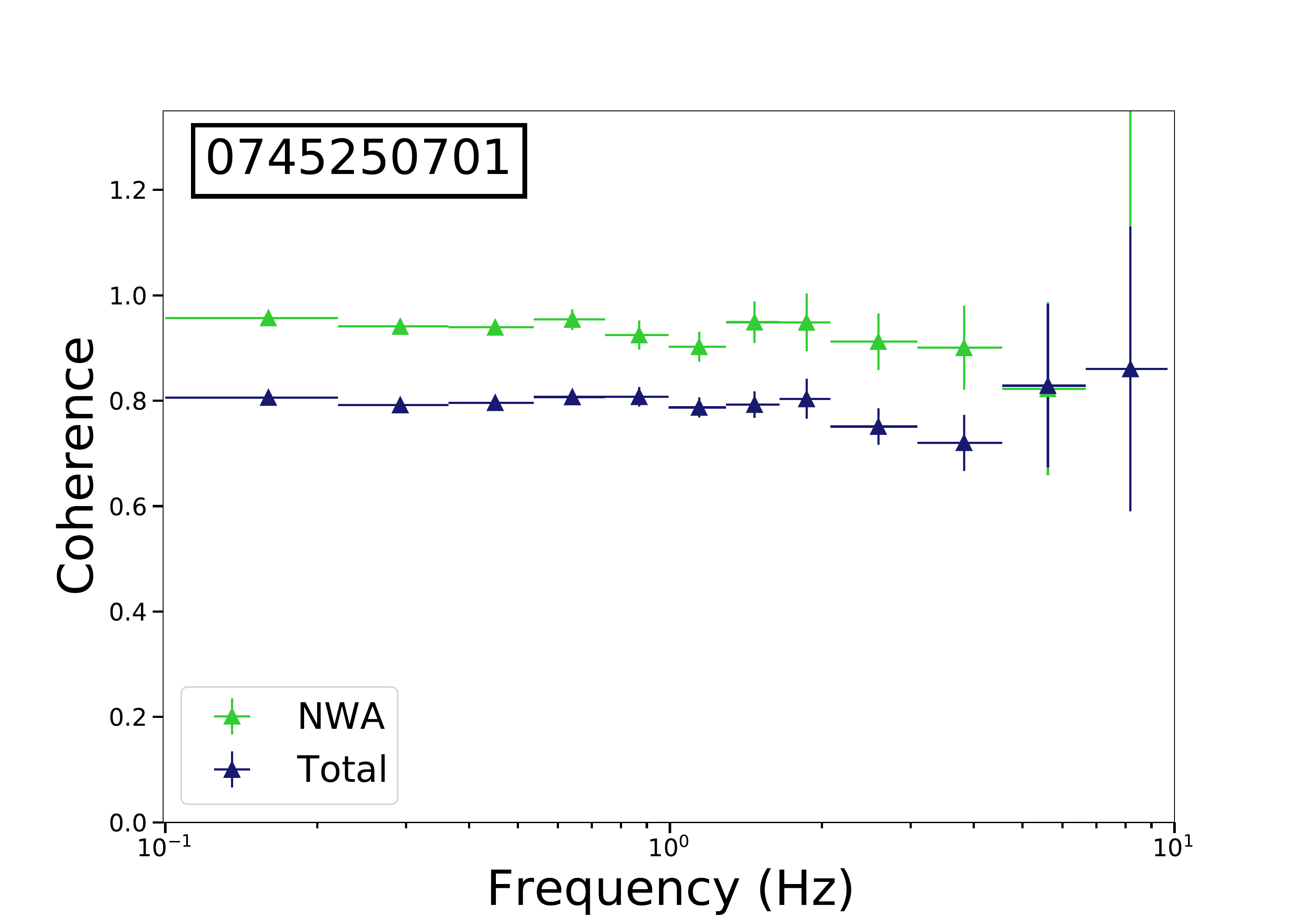}
\caption{Intrinsic coherence between the 0.3--1~keV and the 2--10~keV energy bands for the Total (blue) and NWA (green) datasets. For observation 201, the DIPS dataset (hard colour $\leqslant$~0.8) is shown in magenta}.
\label{cohe}
\end{figure*}

{\label{cohe_freq}}
The coherence measures the degree of linear correlation between two light curves as a function of Fourier frequency \citep[e.g.][]{Vaughan_Nowak1997,Nowak_1999,Uttley_2014}. Fig.~\ref{cohe} shows the intrinsic coherence (computed using equation 8 of \citealt{Vaughan_Nowak1997}) for the Total and the NWA datasets, between the 0.3--1~keV and 2--10~keV light curves. 

Above $\sim$~10~Hz the intrinsic coherence has large uncertainties because of a significant contribution from uncorrelated Poisson noise. Therefore, we decided to limit our analysis to the frequencies less affected by Poisson noise (i.e. 0.1--10~Hz).

We find that the intrinsic coherence of the Total dataset changes significantly among the different observations (ranging between $\sim$~0.6--0.9), and we ascribe this behaviour to the presence of the wind. Indeed, for this dataset, the lowest values of intrinsic coherence are registered during the observations characterized by stronger absorption (observation 201 and 701). 
On the other hand, the intrinsic coherence of the NWA dataset is high ($\sim$~0.95) and consistent with being constant among observations, as commonly observed \citep[e.g.][]{Nowak_1999,Pottschmidt_2003,Grinberg_2014}. In other words, when considering the ``bare'' emission from the X-ray source, there are no indications of significant changes of coherence between variability in the two energy bands.

\subsection{Lag-frequency spectra} 
\begin{figure*}
\centering
\hspace*{-0.4cm} 
\includegraphics[width=0.5\linewidth]{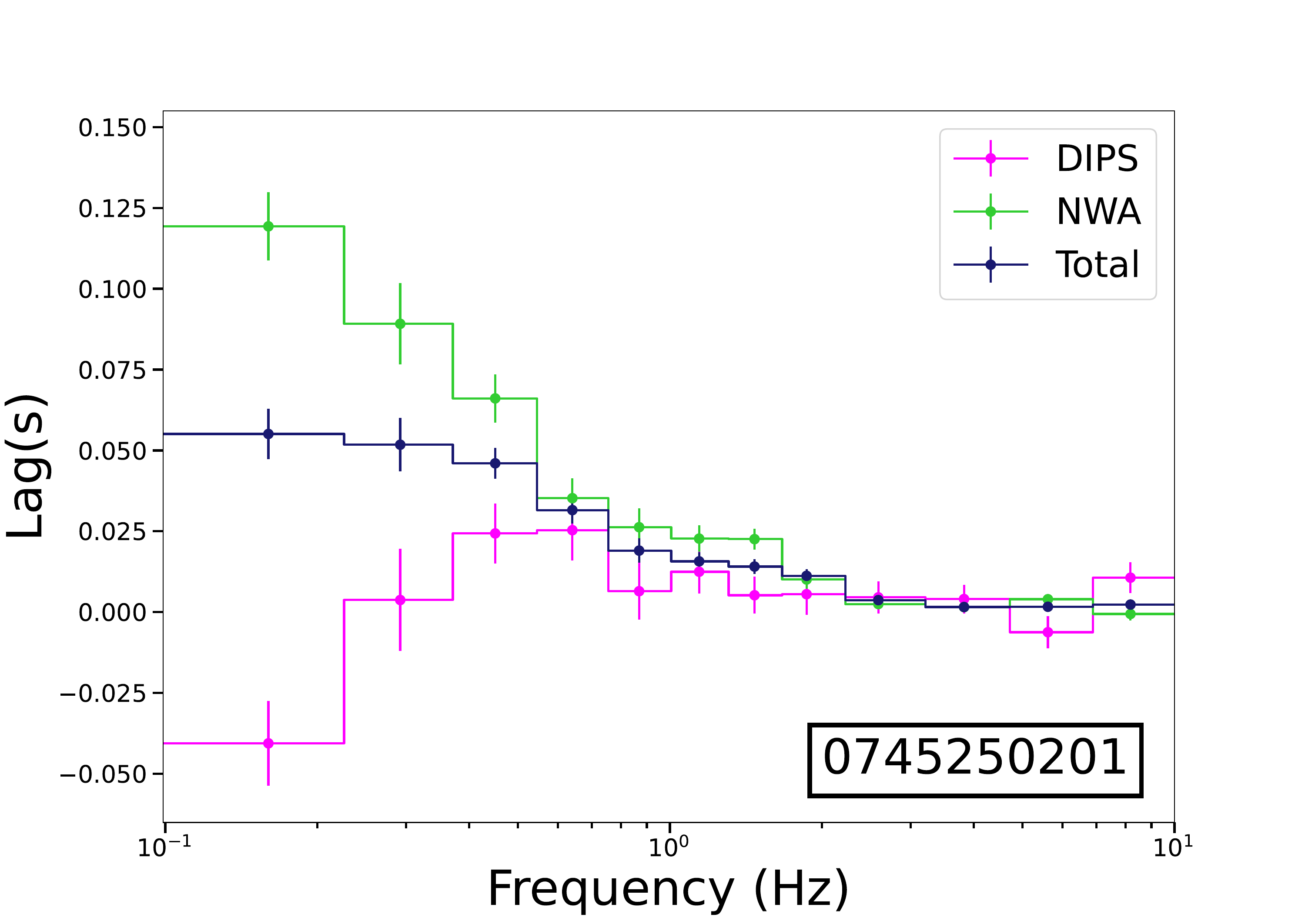}
\includegraphics[width=0.5\linewidth]{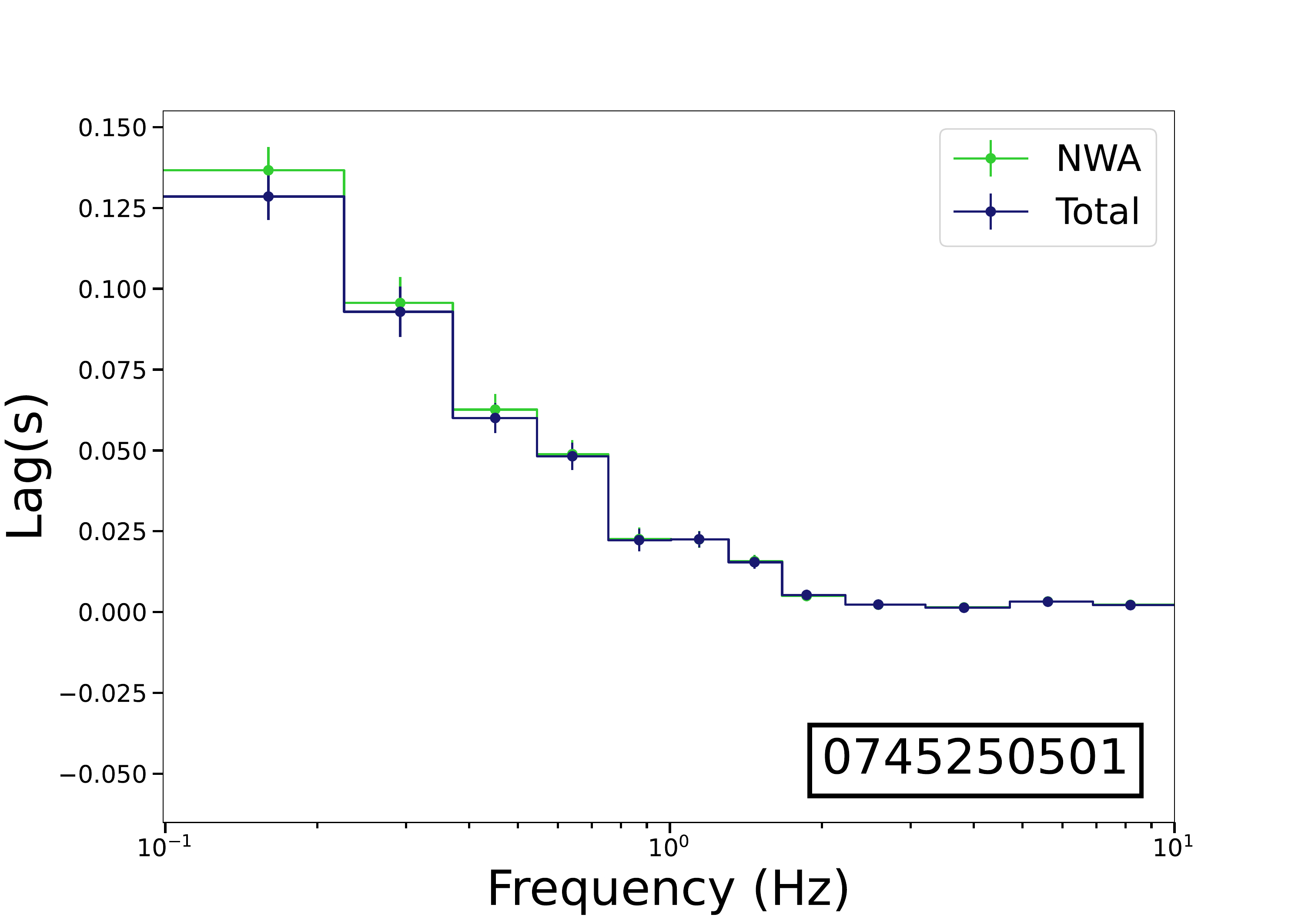}
\hspace*{-0.4cm}
\includegraphics[width=0.5\linewidth]{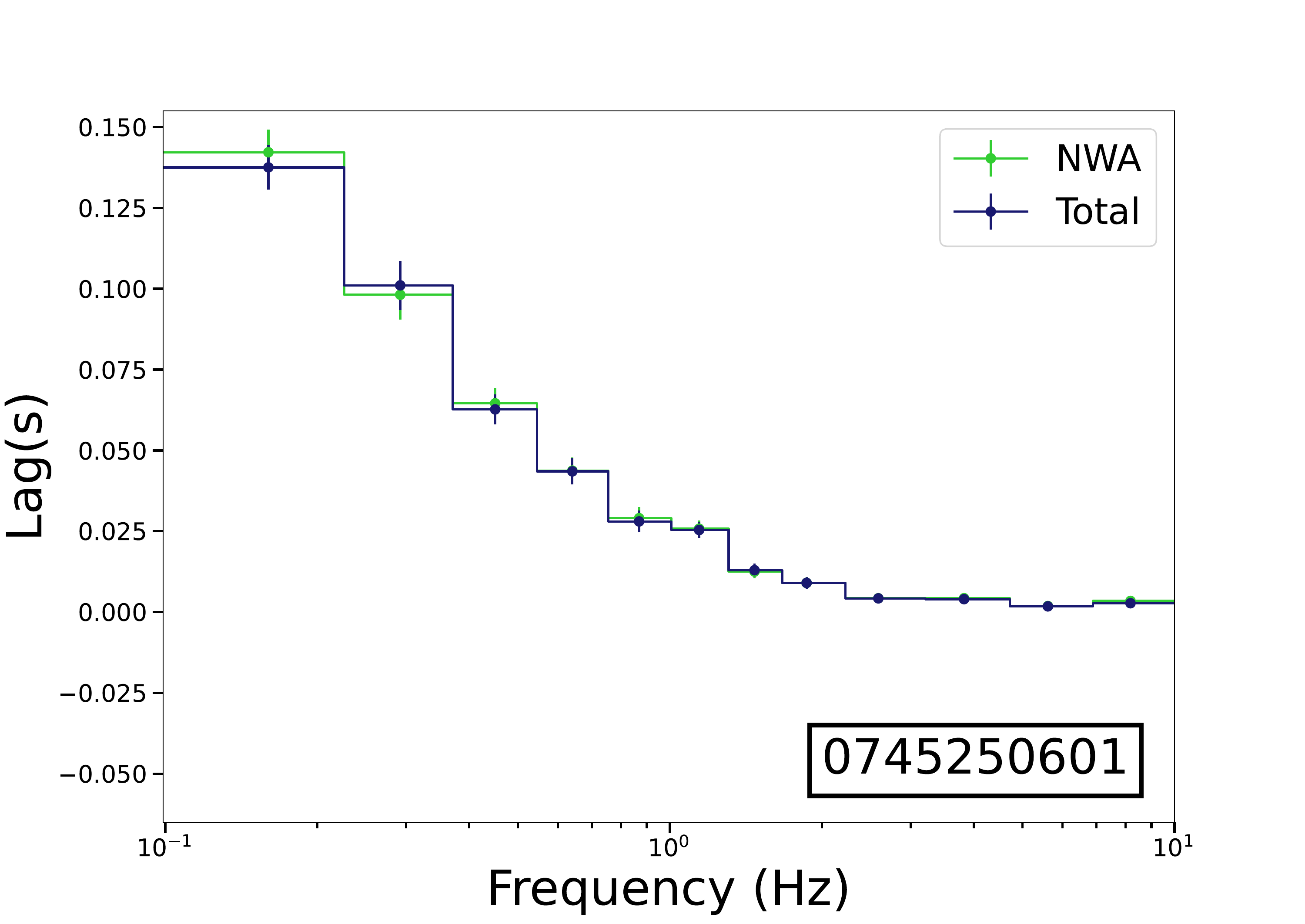}
\includegraphics[width=0.5\linewidth]{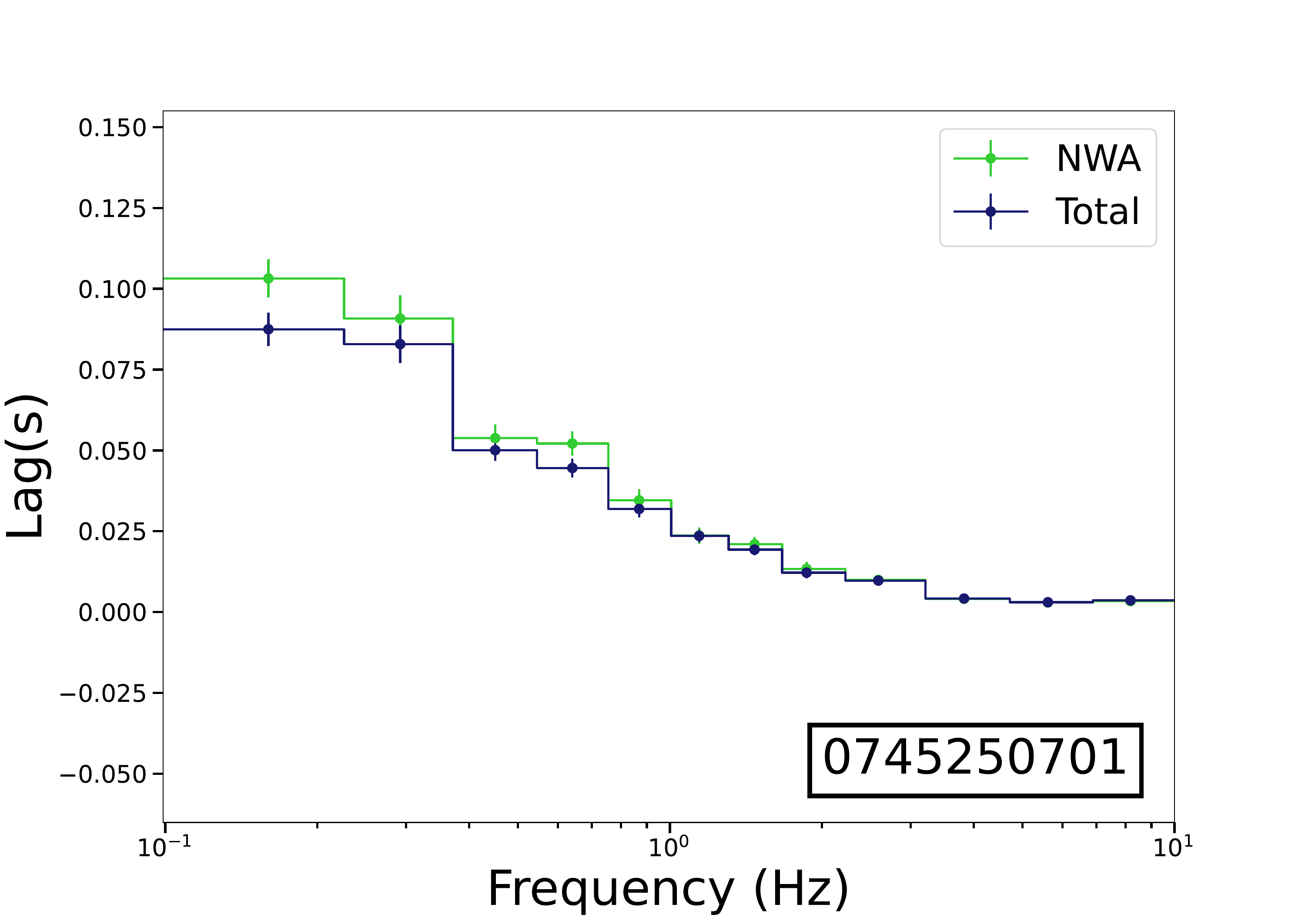}
\caption{Lag-frequency spectra between the 2--10~keV and the 0.3--1~keV energy band for Total (blue) and NWA (green) datasets. In magenta, the DIPS dataset (with a hard colour $\leqslant$~0.8) for observation 201.}
\label{timelag_freq}
\end{figure*}

{\label{lag_freq}}
We then computed the frequency-dependent time lags between the same soft and hard energy bands used to compute the intrinsic coherence in Sect.~\ref{cohe_freq}. The time lags were estimated as $\tau(\nu) = \phi(\nu)/2\pi\nu$, where $\phi(\nu)$ is the phase lag of the average (over the different light curve segments of a single observation) cross-spectrum.
Results are shown in Fig.~\ref{timelag_freq}, for both the Total (blue) and NWA (green) datasets of each observation. 

We observe that the amplitude of the lag is positive at all frequencies, indicating that rapid variability in the 2--10~keV energy band is delayed with respect to the 0.3--1~keV band. Interestingly, there is no evidence of a thermal reverberation lag in the 1--10~Hz range (which would manifest as a negative lag representing a delay of the disc-dominated, 0.3--1~keV band with respect to the Comptonisation-dominated 2--10~keV band).
While hard lags are commonly observed in BHXRBs \citep[e.g.][]{Nowak_1999,Uttley_2011,DeMarco_2015}, we note that the wind affects their amplitude at low frequencies.
Indeed, during observations 201 and (to a lesser degree) 701, the Total datasets display slightly shorter low frequency hard lags than during the other observations. 
We further investigated this issue in Sect.~\ref{DIPS_section} for the most absorbed observation.
After correcting for the effects of wind-absorption (NWA dataset in Fig.~\ref{timelag_freq}), the amplitude of the low frequency hard lags in observations 201 and 701 increases, becoming consistent with the lags measured in the other observations\footnote{The shorter low frequency hard lags for observation 701 could be related to residual wind absorption as mentioned in Sect.~\ref{PS}}. 

\subsection{The X-ray spectral-timing properties of the stellar wind}
\label{DIPS_section}
\begin{figure}
\includegraphics[trim={0cm 0cm 0cm 0cm},width=0.5\textwidth]{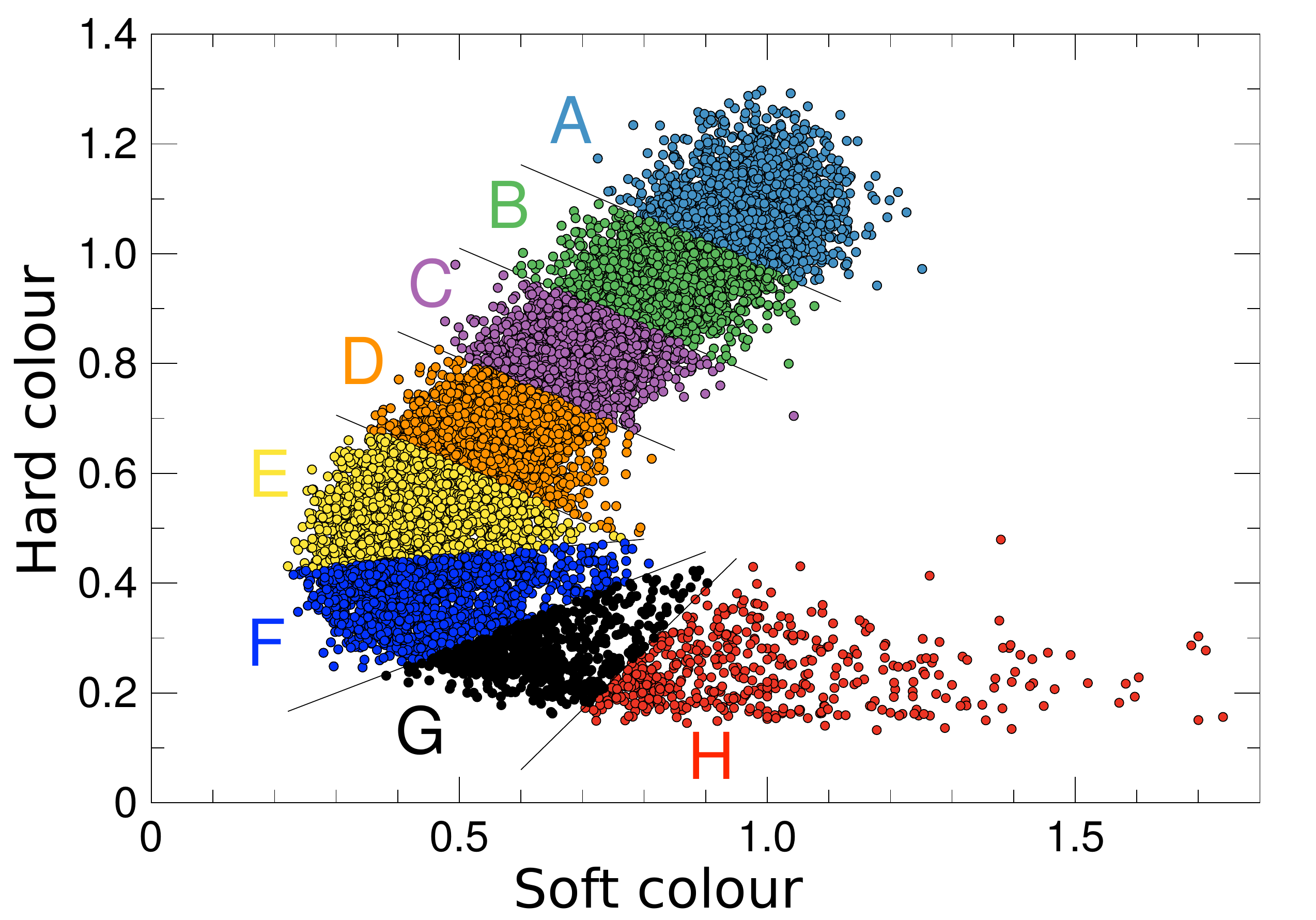}
\caption{The colour-colour diagram of observation 201 showing the regions selected for the study of the PSD as a function of the amount of absorption. From A to H the different regions correspond (for a constant covering factor) to an increasing value of $\mathrm{N_{H,w}}$.}
\label{CCDiagram_T}
\end{figure}

\begin{figure*}
\includegraphics[trim={0cm 0cm 0cm 0cm},width=0.8\textwidth]{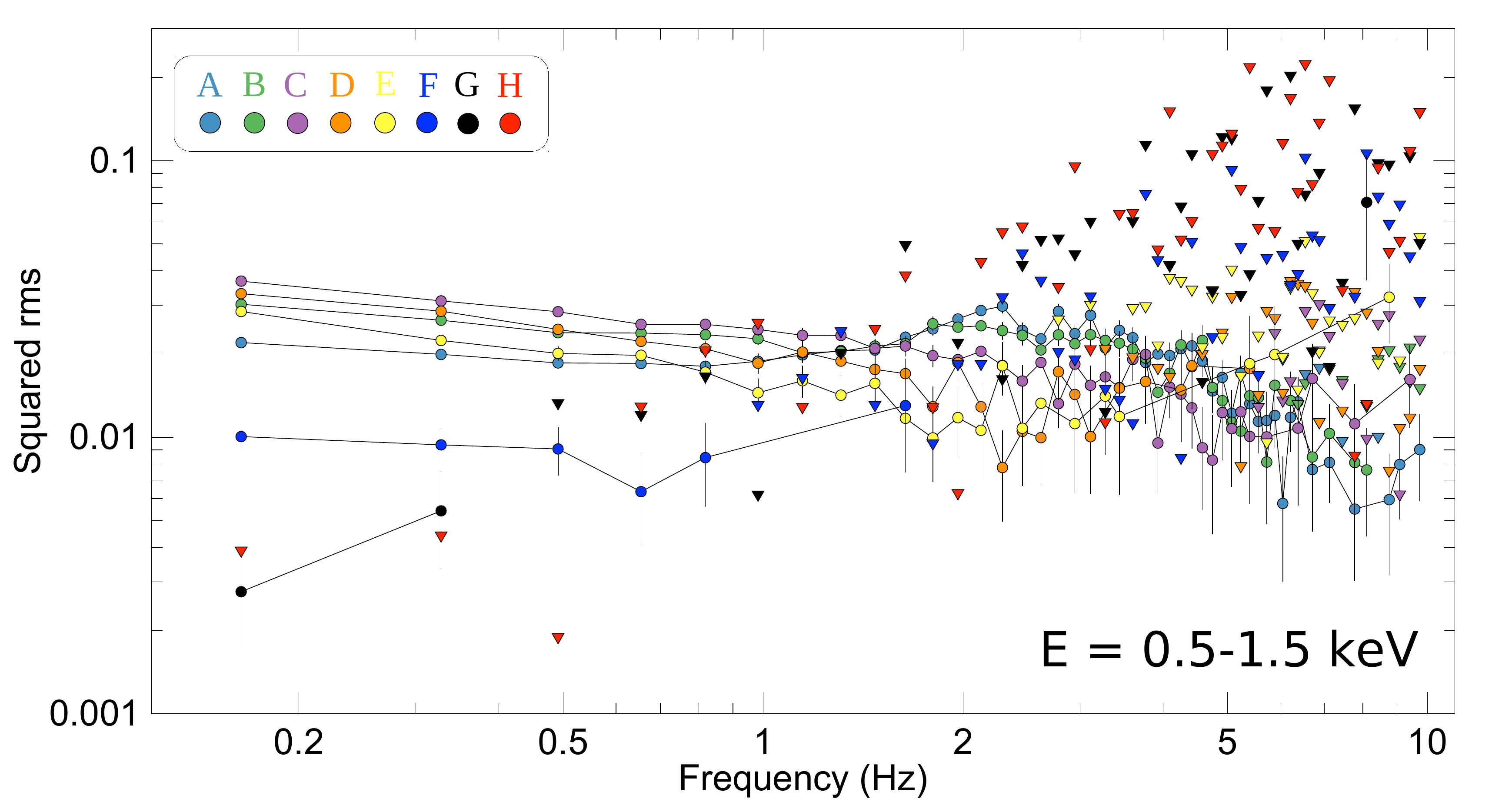}
\includegraphics[trim={0cm 0cm 0cm 0cm},width=0.8\textwidth]{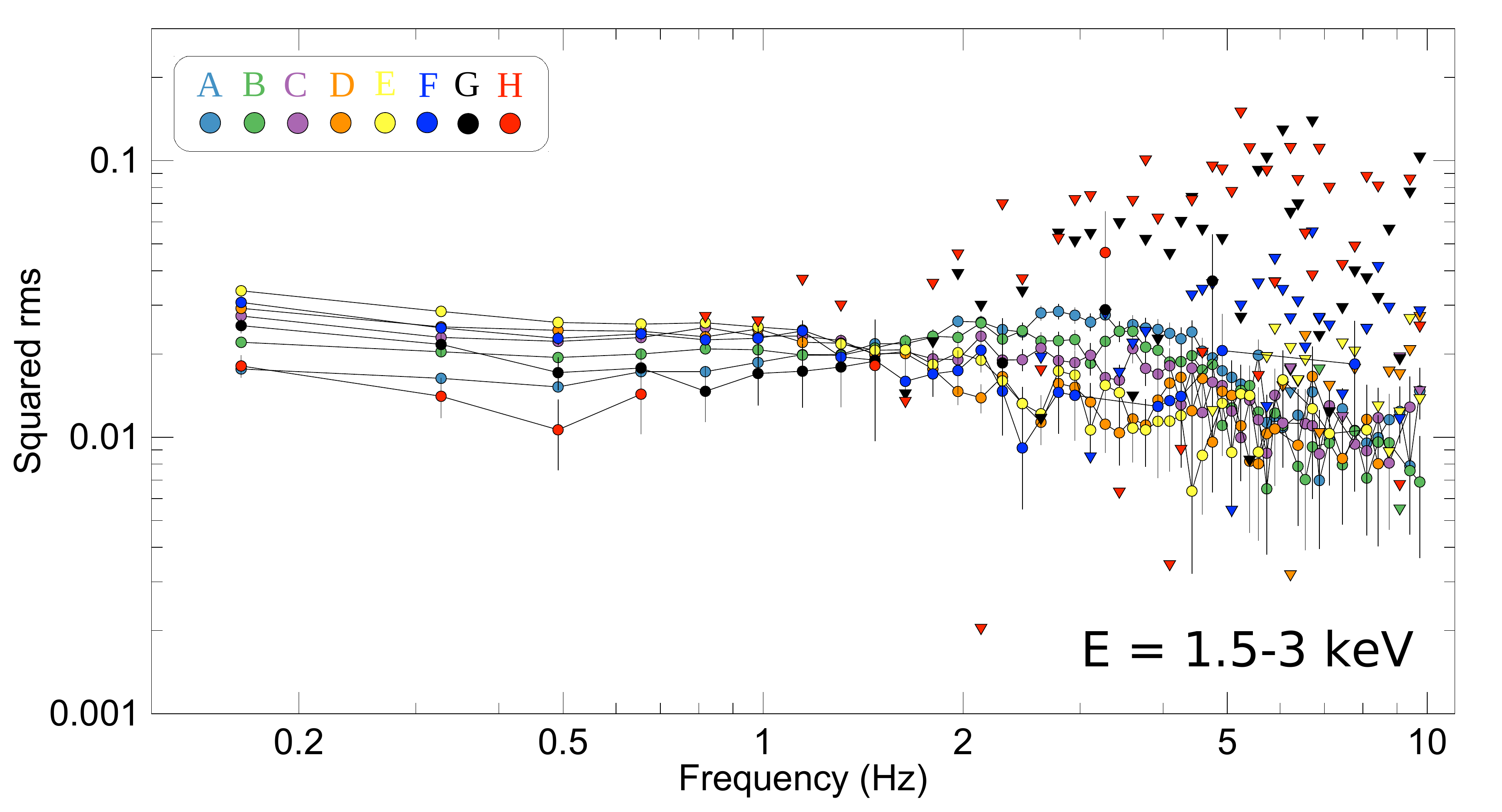}
\includegraphics[trim={0cm 0cm 0cm 0cm},width=0.8\textwidth]{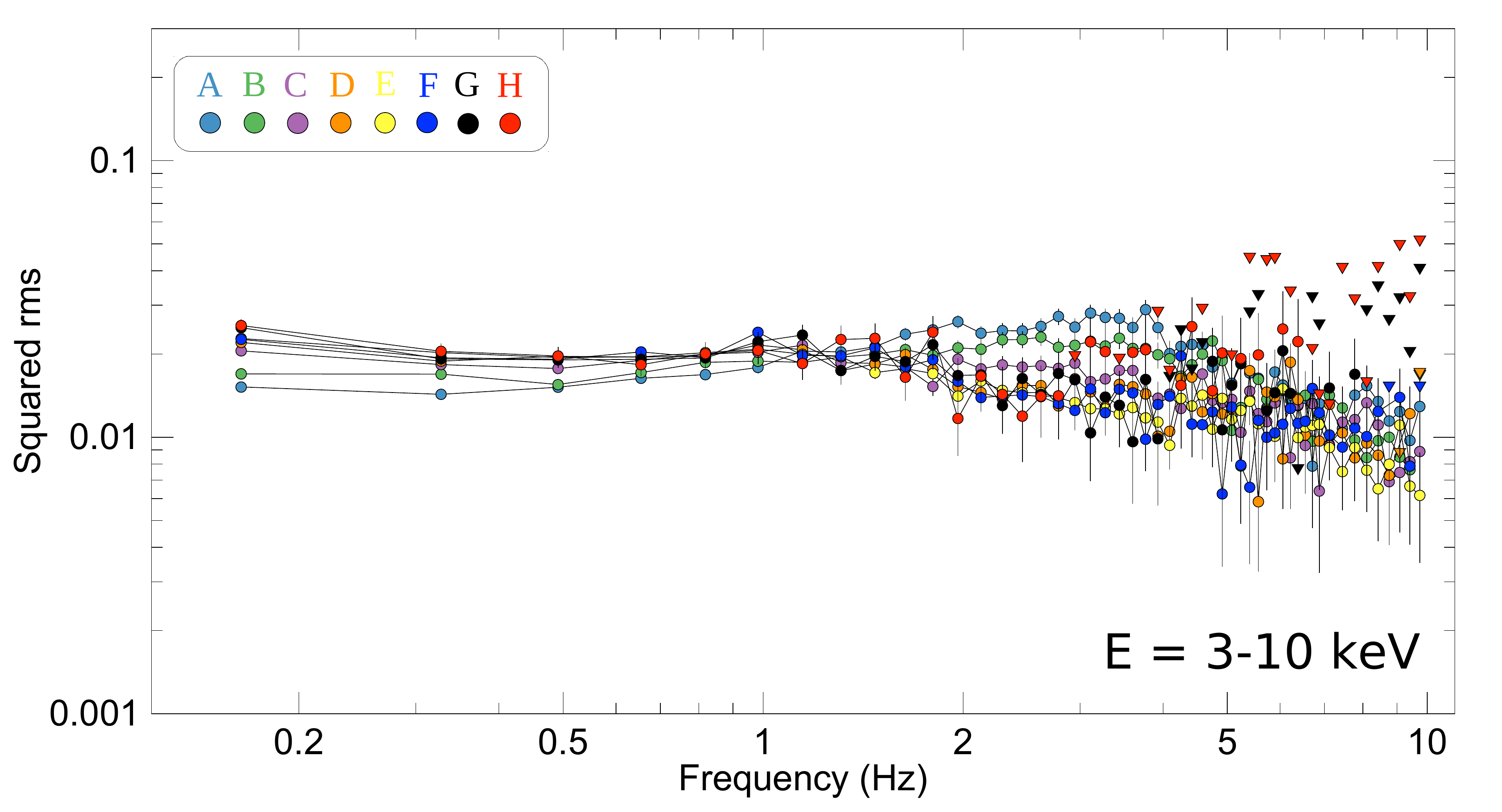}
\caption{PSDs for the selected regions in Fig.~\ref{CCDiagram_T} in the soft (0.5--1.5~keV, upper panel), intermediate (1.5--3~keV, middle panel) and hard (3--10~keV, lower panel) bands. The triangles represent upper limits (at $3\sigma$ confidence level).} 
\label{PS_T}
\end{figure*}

Given the observed differences between the Total and the NWA datasets, we investigated in more detail the spectral-timing signatures of the X-ray variability induced by the stellar wind. We did so by analysing time intervals dominated by wind absorption, thus corresponding to the deeper phases of each X-ray dipping event (Fig.~\ref{HIDiagram}). To this aim we focused on observation 201 (which covers the first passage at superior conjunction), as this displays the largest number of absorption dips in the light curve and the densest track in the colour-colour diagram of the source (Figs.~\ref{HIDiagram} and \ref{fig:CCDiagram}).
We first studied the dependence of the measured fractional rms on the amount of absorbing material along the line of sight. We selected 8 different regions of soft and hard colours which, for a given covering factor (Appendix \ref{APP_CCDiagram_model_data}), would correspond to increasing values (from A to H in Fig.~\ref{CCDiagram_T}) of the column density $\mathrm{N_{H,w}}$. 
Using the same procedure 
as in Sect.~\ref{PS}, we computed the average PSD of the data within each selected regions. The PSDs are extracted in the soft, intermediate and hard energy bands used in Sect.~\ref{selezione_NWA}, and are displayed in Fig.~\ref{PS_T}.
At both low and high frequencies we can observe significant differences as a function of the amount of intervening gas. In particular, we register a gradual increase of low frequency variability and a decrease of high frequency variability power. Above a certain value of $\mathrm{N_{H,w}}$ (regions D to H of the colour-colour diagram), in the soft and in the intermediate bands (Fig.~\ref{PS_T}, upper and middle panel) the variability power starts to be suppressed also at lower frequencies. In other words, at the bottom of the dip, soft band variability is significantly damped on all sampled timescales, while this suppression appears to occur only above an increasingly higher frequency as the energy increases. 

\begin{figure}
\includegraphics[trim={0cm 0cm 0cm 0cm},width=0.48\textwidth]{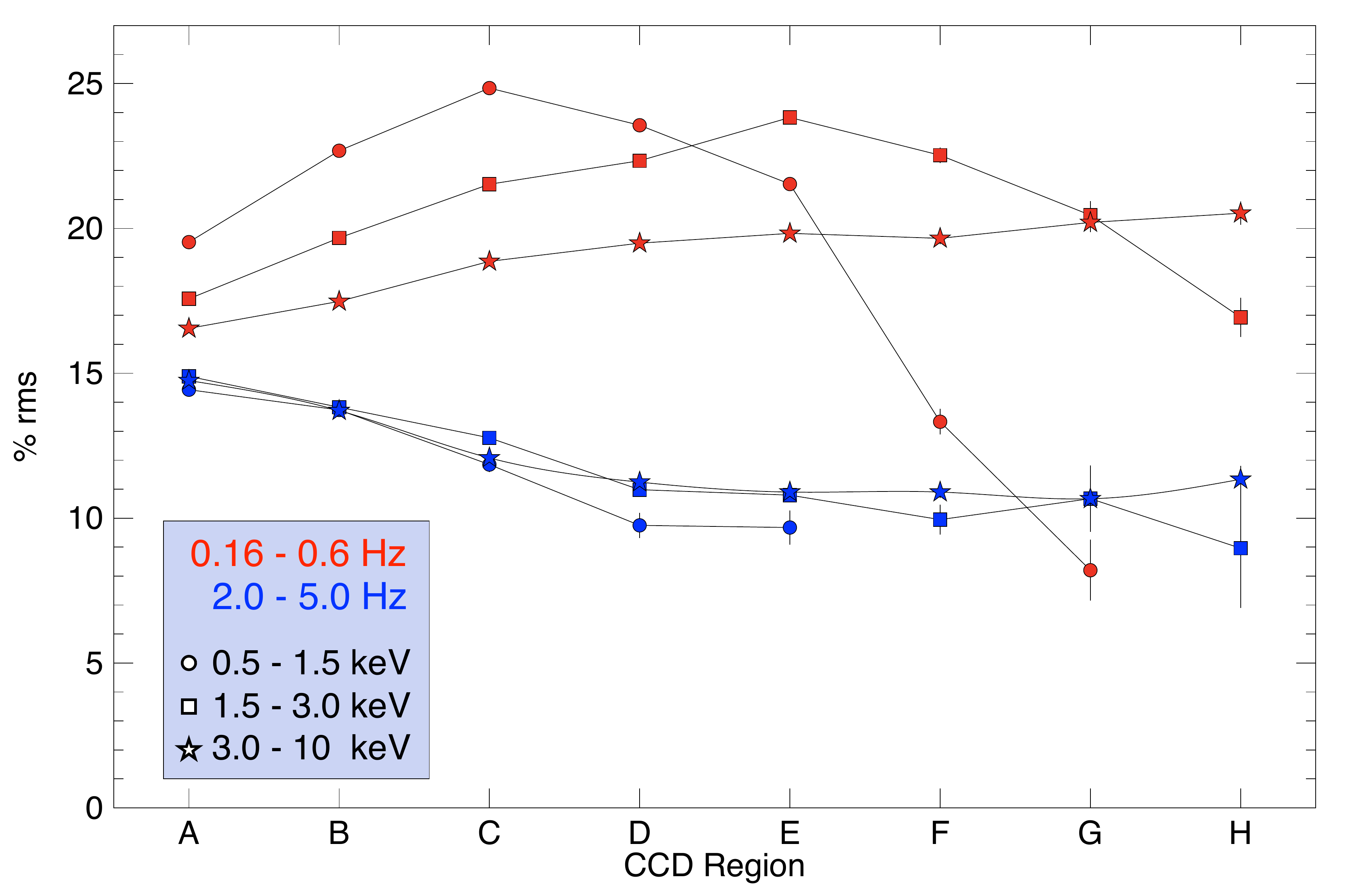}
\caption{Fractional rms as a function of increasing wind absorption (from region A to region H).}
\label{frac_var_T}
\end{figure}

In order to better visualise these trends, we integrated the PSDs over the 0.16--0.6~Hz and the 2--5~Hz frequency intervals. The resulting fractional rms in each frequency range is plotted separately (see Fig.~\ref{frac_var_T}) for each energy band as a function of the selected region of the colour-colour diagram. 

The 0.16--0.6~Hz fractional rms initially increases up to $\sim$~23--25 percent. Then it starts decreasing, reaching values of $\sim$~10--17 percent. This change in the overall trend occurs in regions of the colour-colour diagram characterised by higher values of $\mathrm{N_{H,w}}$. In the softest band the trend reversal occurs earlier in the colour-colour diagram track (i.e. at higher values of the hard colour) than seen in higher energy bands. At soft energies, the most affected by absorption, we also register the highest/lowest values of maximum/minimum fractional rms. While the hardest energy band does not show evidence of a drop of 0.16--0.6~Hz fractional rms, an increasing trend can still be clearly observed, suggesting absorption affects also this region of the spectrum. 

On the other hand, the 2--5~Hz fractional rms steadily decreases (from $\sim$~15 to $\sim$~12 percent) in all bands, eventually reaching a plateau (between regions D and H). 

The increase of fractional rms, observed at low frequencies in some regions of the colour-colour diagram and at different levels for different energy bands, can be explained in terms of variations of the column density of the absorbing gas crossing the line of sight to the X-ray source. However, the suppression of variability that occurs in the most absorbed phases/energy bands, and at the highest sampled frequencies requires the presence of additional scattering components (see discussion in Sect.~\ref{discussion}).

Finally, we measured again the coherence and the lags between the 0.3--1~keV and the 2--10~keV energy bands (as in Sects. \ref{cohe_freq} and \ref{lag_freq}), this time using a dataset which selects only intervals strongly affected by the wind (i.e. with a hard colour $\leqslant$~0.8). This dataset (hereafter referred to as ``DIPS'') roughly corresponds to the regions D to H in Fig.~\ref{CCDiagram_T}. The results are overplotted in Figs. \ref{cohe} and \ref{timelag_freq}. A comparison with the Total and NWA datasets clearly demonstrates that wind absorption is the main cause of the lower coherence and shorter hard lags measured in the Total dataset during each passage at superior conjunction. Indeed, the X-ray variability associated with the stellar wind is characterised by a low intrinsic coherence ($\lesssim$~0.5, and dropping to zero above 2~Hz) and a negative (soft) lag of a few tens of msec.

\section{Discussion}\label{discussion}

Detailed studies of the stellar wind in Cyg X-1 have been mostly focused on characterizing the way the wind influences the spectral properties of the system, and generally limited to $E\gtrsim 1$~keV \citep[e.g.][]{Grinberg_2015, Hirsch_2019}. On the other hand, studies of the X-ray timing properties of Cyg X-1 have been mostly focused on the X-ray source, neglecting the effects of the wind \citep[e.g.][]{Pottschmidt_2003,Axelsson_Done_2018,Mahmoud_2018_A,Mahmoud_2018_B}.

In this paper we found that the presence of a stellar wind significantly affects the X-ray variability power of the source, as well as the amount of coherence and the amplitude of the lags between variable emission in the soft (0.3--1~keV) and hard (2--10~keV) energy bands. The observed behaviour changes significantly as a function of the orbital phase, due to the orbital modulation of wind absorption.

\subsection{Effects of the wind on the X-ray variability power}
\label{diss_PS}

We found that the stellar wind influences the observed X-ray variability power of the source in a complex way, that depends on both energy and timescale (Sects. \ref{PS} and \ref{DIPS_section}). These results are discussed below considering the low and high frequency behaviour separately.\\

\subsubsection{X-ray variability at low frequencies ($\lesssim$~1~Hz)}

The variable X-ray absorption associated with the stellar wind has the effect of increasing the long timescale/low frequency fractional rms of the source (Figs.~\ref{psd}, \ref{PS_T}, and red 
markers in Fig.~\ref{frac_var_T}). This wind-induced, excess X-ray variability is observed in all the energy bands considered, including the highest energies (3--10~keV), in line with spectral studies that showed that the stellar wind can modify the broad band continuum up to very high energies \cite[$\sim$~10~keV, e.g.][]{Grinberg_2015}. 

By considering a clumpy wind model, \cite{Grinberg_2015} investigated the variability induced by clumps of variable size and porosity (a measure of the mean free path among clumps), and constant mass, as they cross our line of sight. The main inferred effect of these transits is an increase of the observed variability power due to the variations of the $\mathrm{N_{H,w}}$ of the intervening structured wind. According to this study, excess X-ray variability would be observed only when averaging over timescales sufficiently short so as to catch the passage of a single/small number of clumps (at lower frequencies/longer timescales wind-induced variability would result washed out as a consequence of averaging over a large number of transits).
Although this model may represent a simplification of the real conditions, it clearly shows how variations of the $\mathrm{N_{H,w}}$ of the wind can produce an increase of variability power as observed in the data analysed here (Figs. \ref{psd}, \ref{PS_T}, and \ref{frac_var_T}). This would imply that the timescales sampled in our study ($\sim$~1--10~s) are short enough to catch the passage of single clumps.
\cite{Grinberg_2015} also note that below/above a certain timescale/temporal frequency the wind does not contribute to increase the level of X-ray variability further. This timescale roughly corresponds to the time needed for a typical clump to transit the line of sight, and thus can be used (knowing the wind velocity) to estimate the typical size of the clumps. We can obtain an estimate of the typical size of the clumps responsible for the observed excess variability in Cyg X-1 by considering the maximum frequency, $\nu_{\mathrm{max}}$, at which the variability power in the Total PSD is observed to be greater or equal to the variability power in the NWA PSD. Assuming a wind terminal velocity of $v_{\infty}=2400\ {\rm km\ s^{-1}}$ \citep{Grinberg_2015}, the resulting radial size of the clumps would be of the order of $l \sim$~0.5--1.5 $\times 10^{-4} R_{\ast}$ (we note that the larger values of $\sim$~0.02--0.2 $R_{\ast}$ previously found by \citealt{Hirsch_2019} are due to the larger minimum timescale sampled in their analysis, i.e. 25.5 s).
This range accounts for the different values of $\nu_{\mathrm{max}}$ observed in the different observations. We note however, that while this represents an order of magnitude estimate, the clumps likely have a broad distribution of sizes. In particular, smaller clumps can give a significant contribution to the observed high frequency X-ray variability.

When considering the most absorbed stages (which correspond to lower values of hard colour in the colour-colour diagram, Fig.~\ref{CCDiagram_T}, and for a given covering factor, to higher $\mathrm{N_{H,w}}$, Fig.~\ref{APP_CCDiagram_model_data}), 
we found that the low-frequency X-ray variability power in softer bands starts to be significantly suppressed (Fig.~\ref{frac_var_T}). Large scale reprocessing/fluorescent scattering of soft band photons in the stellar wind is a plausible mechanism to explain the decrease of low frequency variability power during the dips. Indeed, light travel time differences across this extended medium strongly dilute the intrinsic flux variability of the X-ray source. This constant or slowly variable extended emission component from the wind would dominate the soft X-ray band when the direct continuum is highly absorbed, resulting in a net decrease of fractional rms at these energies. We note that a similar effect might be produced by scattering off dust layers in the ISM (known to produce a soft spectrum due to the energy dependence of the dust scattering cross-section, \citealt{Predehl_Schmidt_1995,Xiang_2011,Ling_2009}). However, this seems less plausible given the factor of $\sim$~2 decrease in fractional rms observed during the dips in the softest energy band (Fig. \ref{frac_var_T}), which implies quite a large scattering fraction.

\subsubsection{X-ray variability at high frequencies ($\gtrsim$~1~Hz)} 
The observations during passage at superior conjunction (observations 201 and 701) are characterized by a strong suppression of the high frequency variability hump, a characteristic feature in the power spectrum of Cyg X-1. This can be seen by comparing the Total (as well as the DIPS for observation 201) datasets with the NWA datasets (Figs. \ref{psd} and \ref{fig:APP_701}). Such a suppression is observed at all energies, and according to Fig. \ref{frac_var_T}, it starts occurring at relatively low levels of wind absorption.
Nonetheless, this behaviour cannot be ascribed to absorption in the stellar wind, since variable absorption would result in an increase of fractional rms, while constant absorption would equally affect the variable and constant flux components, thus it would not produce any change in the fractional rms.

Here, we consider the possibility that the suppression of the high frequency variability power is due to scattering, thus smearing out the fastest variability, and infer the optical depth of the scattering medium. The timescales affected by scattering will be those shorter than the light crossing time of the scattering medium, thus corresponding to the highest temporal frequencies in the power spectrum. The temporal frequency where this dampening becomes significant is expected to decrease as the optical depth $\tau_s$ increases \citep[e.g.][]{Zdziarski_2010}.

\begin{figure}
\centering
\includegraphics[trim={0cm 0cm 2cm 1cm},width=1\linewidth]{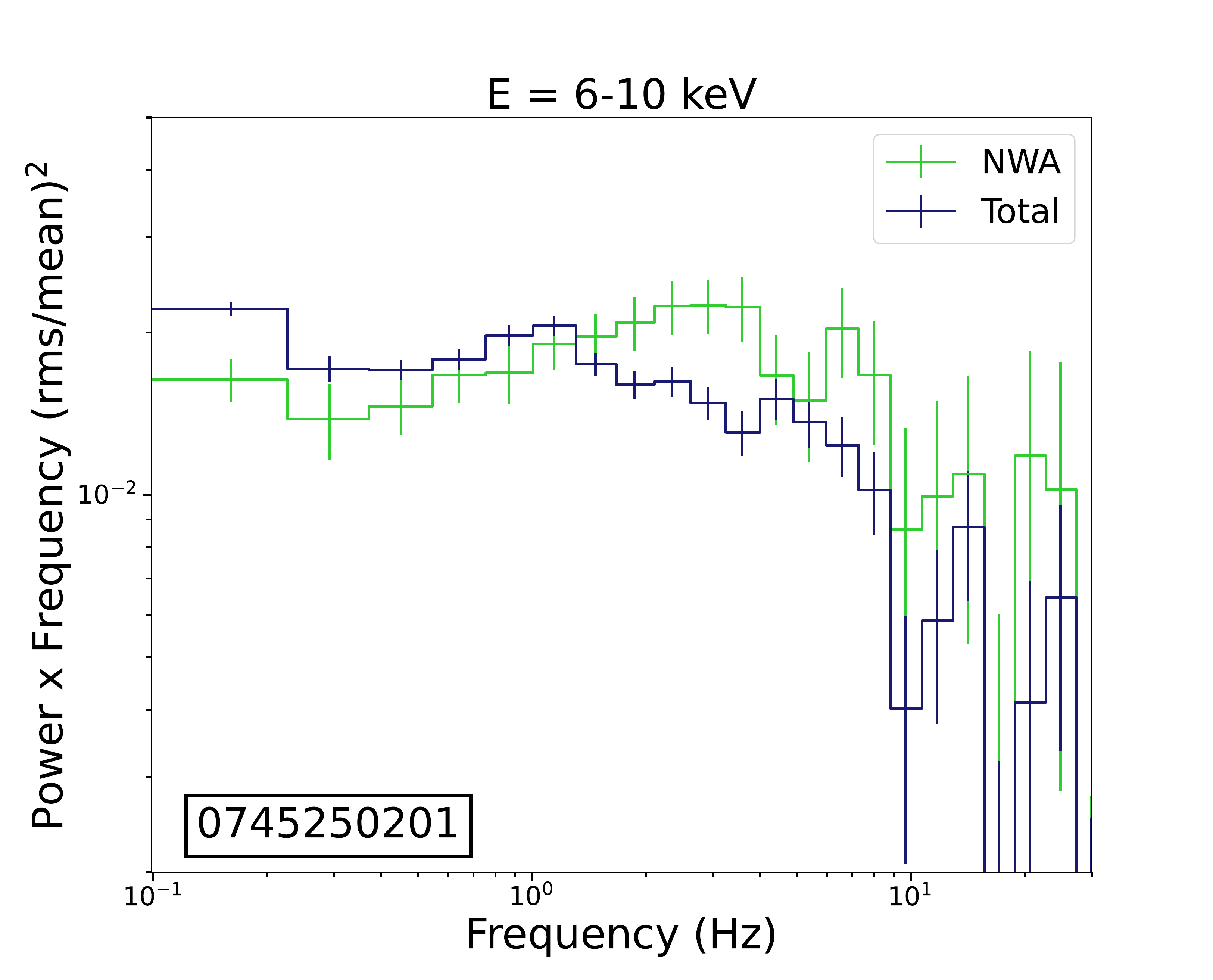}
\caption{PSD of observation 201 for the Total (blue) and NWA (green) datasets in the 6--10~keV energy band.}
\label{PS_6_10keV}
\end{figure}

\begin{figure}
\includegraphics[trim={0cm 0cm 0cm 0cm},width=0.5\textwidth]{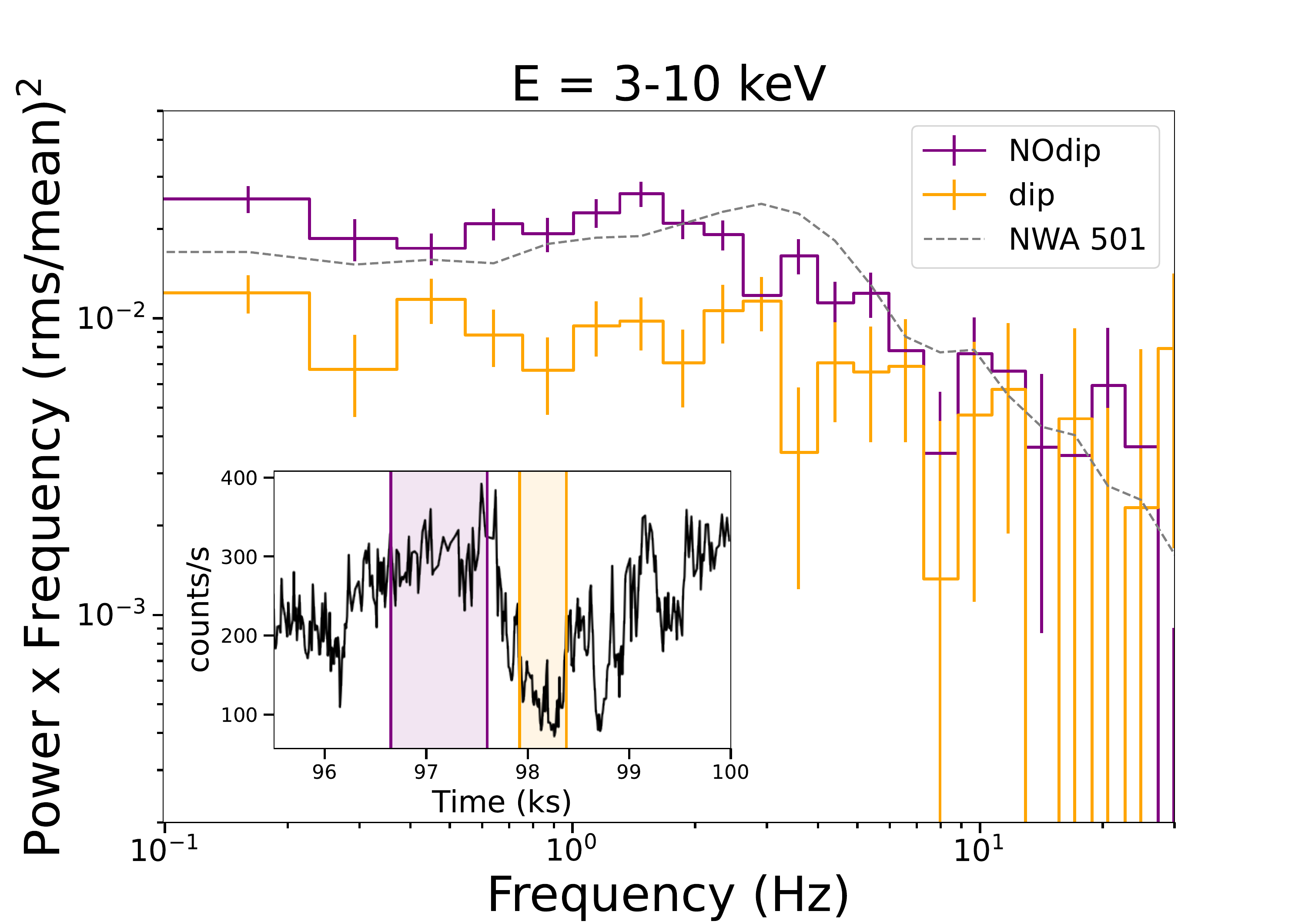}
\caption{PSDs extracted from two adjacent time intervals (see inset) during observation 201 near superior conjunction, respectively characterized by low (`NOdip'', purple) and strong (`dip'', orange) absorption. The PSDs are extracted in the 0.3--10~keV energy band. Overplotted for comparison is the NWA PSD of observation 501 near inferior conjunction (dashed line), which shows the intrinsic shape of the PSD of Cyg X-1 (i.e. free from absorption and possible scattering effects at superior conjunction).}
\label{PS_LC}
\end{figure}

Such a behaviour has been previously observed in a number of sources \citep[e.g.][]{Belloni_1991,Berger_vanderKliss_1994}. Interestingly, in the case of the BHXRB Cyg X-3 the analysis of the PSD required the presence of a gas optically thicker than the wind \citep[][]{Zdziarski_2010}. Using equation 5 of \cite{Zdziarski_2010} we estimate that the drop observed in Cyg X-1 (at $\nu\sim$~5--10~Hz) would require $\tau_s \sim$~0.5--1 for a scattering region of the size of $R\sim$~3$\times 10^9$~cm, as inferred for Cyg X-3. The need for such a large $\tau$ is also confirmed by the fact that the suppression of high frequency variability in the PSD of observation 201 is still observed at very high energies (see Fig.~\ref{PS_6_10keV}, reporting the PSD of the Total and NWA datasets of observation 201, extracted in the $E=$~6--10~keV energy band). These results support the hypothesis of the presence of an optically thick gas intercepting the line-of-sight to the X-ray source during superior conjunction. However, previous analyses suggest $\tau_w \ll 1$ for the stellar wind in Cyg X-1 \citep{Zdziarski_2012}, remarkably lower than required from our estimates. Therefore, as previously seen in Cyg X-3, a gas optically thicker than its stellar wind is needed to explain the drop of high frequency variability seen in Cyg X-1. 
This optically thicker gas may be associated with an accretion bulge, possibly formed by collision of the stellar wind with the edge of the disc \citep[e.g.][]{Poutanen_2008,Zdziarski_2010}. 
To further test this hypothesis, we extracted PSDs of observation 201 from two adjacent time intervals selected near superior conjunction, one characterized by low absorption and the other dominated by strong dips (as shown in the inset of Fig.~\ref{PS_LC}).
The high frequency humps are dampened in both PSDs (Fig.~\ref{PS_LC}, see the comparison with the NWA dataset of observation 501, which shows the shape of the PSD, far from superior conjunction), i.e. both inside and outside the strongest absorption dips, thus excluding denser wind clumps as the origin of the variability drop, and supporting the hypothesis of the presence of a scattering bulge.

Finally, we note that the high frequency variability power is not completely suppressed. Indeed, studying observation 201 in more detail, we observe that the decrease of high frequency fractional rms is followed by a plateau during the most absorbed stages (Fig.~\ref{frac_var_T}). This high frequency variability may be due to intrinsic source variability, from photons that reach the observer without being scattered, or to residual contribution from variable absorption. For example, considering the clumpy wind scenario discussed above, wind clumps producing excess variability at low frequencies as they cross our line of sight, will likely have a distribution of sizes. This implies that smaller clumps can exist which would contribute to the high frequency X-ray variability of the source.

\subsection{Effects of the wind on X-ray coherence and lags}

We found that another important effect associated with the presence of the stellar wind is a decrease of intrinsic coherence between the soft (0.3--1~keV) and the hard (2--10~keV) energy bands (Fig.~\ref{cohe}) on all time scales. 
On the other hand, in agreement with previous studies, the intrinsic coherence of the source is close to 1 when time intervals free from wind absorption are selected (NWA in Fig.~\ref{cohe}).
In addition, we showed that the amplitude of the hard X-ray lags, commonly observed in Cyg X-1 and a characteristic feature of BHXRBs \citep[e.g.][]{Uttley_2011,DeMarco_2015}, results reduced when the source is at superior conjunction. This effect is particularly strong during observation 201 (Fig.~\ref{timelag_freq}), possibly due to denser clustering of absorption events (Fig.~\ref{HIDiagram}).

The observed loss of coherence requires the emergence of one (or more) non-linear variability components during the most absorbed phases of the orbit, and it is clearly a direct or indirect consequence of the presence of the stellar wind. 
Indeed, when considering only the most absorbed stages of observation 201 (DIPS in Fig.~\ref{cohe}, upper left panel) low coherence is observed at all frequencies (the average low frequency coherence is $\sim$~0.5, a factor of $\sim$~2 lower than in the NWA dataset). 
Given the presence of the wind absorbing medium, the most plausible way to explain the observed loss of coherence between highly absorbed soft bands and primary continuum-dominated hard bands is the non-linearity of absorption variability. This may be due either to motions of clouds that are optically thicker in soft than in hard bands, or to the non-linear response of the absorbing gas to variations of hard X-ray irradiation. We note that a similar drop of coherence ascribable to intervening absorption structures was reported in the Seyfert 1 galaxy NGC 3783 \cite{DeMarco_2020}. Simulations showed that changes in the photo-ionisation state of the wind, as a consequence of variability of the irradiating X-ray flux, could explain the observed decrease of coherence during obscured states in that source (see also the recent results from \citealt[][]{Juranova_2022}).

Interestingly, the DIPS dataset of observation 201, reveals the presence of an additional, low frequency ($\lesssim$~0.2~Hz) soft (negative) lag component (Fig.~\ref{timelag_freq}, upper left panel), which has the net effect of reducing the amplitude of the observed low frequency hard X-ray lags when the wind-absorbed data are not filtered out (Total in Fig.~\ref{timelag_freq}, upper left panel). This lag is likely the result of large scale scattering off the wind, becoming dominant when the direct continuum is blocked due to strong line-of-sight absorption. Another possibility is that the lag is due to recombination delays \citep{Silva_2016}, thus scaling inversely with the density of the absorbing gas \citep[][]{Nicastro_1999,Behar_2003,Krongold_2007,Kaastra_2012}. However, a detailed modelisation with appropriate spectral-timing models and a higher energy-resolution study of the low frequency soft lag associated with the DIPS dataset would be needed to confirm this hypothesis.

\section{Conclusions}\label{conclusions}

In this paper, we studied how the stellar wind influences the observed X-ray spectral-timing properties of Cyg X-1, focusing on the short timescales ($\lesssim$~10~s), which map the innermost regions of the accretion flow. To this aim we made use of the XMM-Newton data from the CHOCBOX monitoring campaign, which allowed us to extend our analysis down to soft X-ray energies ($E\gtrsim 0.3$), where wind absorption is stronger. The monitoring covers two consecutive passages at superior conjunction (i.e. $\phi_{\mathrm{orb}}=0$). This phase of the orbit is characterized by intense clustering of strong absorption dips caused by the stellar wind, Fig.~\ref{HIDiagram} \citep[e.g.][]{Balucinska_2000,Feng_Cui_2002,Ibragimov_2005,Wilms_2006,Poutanen_2008,Boroson_Vrtilek_2010,Grinberg_2015}.

Our main findings can be summarised as follows. 

\begin{itemize}
\item The most absorbed orbital phases are characterized by an increase of the fractional variability at low frequencies ($\lesssim$~1~Hz), and a suppression at high frequencies ($\gtrsim$~1~Hz). As a consequence, the double-hump PSD shape typical of the hard state of Cyg X-1 results smoothed out when the source is at superior conjunction.
In the least absorbed orbital phases (away from superior conjunction) only the increase of low frequency fractional rms is observed, while the high-frequency PSD hump is not affected.\\

\item We ascribe the increase of low frequency fractional variability power due to variations of the column density of the intervening clouds (e.g. as a consequence of relative motions of the wind clumps). The observed timescales correspond to an average radial size for the clumps of $l \sim$~0.5--1.5$\times 10^{-4} R_{\ast}$ 
(assuming a wind terminal speed of $2400\ {\mathrm{km}\ \mathrm{s}^{-1}}$). \\

\item The suppression of high frequency variability power in the observations at superior conjunction requires the presence of an optically thicker (than the gas in the wind) medium. We associate this component with an accretion bulge, possibly formed by collision of the stellar wind with the edge of the disc. The presence of this component was independently proposed in the past to explain some timing properties of Cyg X-1 and Cyg X-3 \citep[][]{Poutanen_2008,Zdziarski_2010}.\\

\item The most absorbed stages of the orbit are characterized by a decrease of intrinsic coherence between soft (absorption-dominated) and hard (primary continuum-dominated) bands. We ascribe the loss of coherence to non-linear variability of the absorbing gas. In addition, the observation most affected by absorption dips shows the emergence of a long, low frequency soft lag, which contributes to reduce the amplitude of the hard X-ray lags intrinsic to the broad band continuum. A plausible explanation for this lag is large scale reprocessing or recombination within the wind.

\end{itemize}

\section*{Acknowledgements}
\addcontentsline{toc}{section}{Acknowledgements}
This work is based on observations obtained with XMM-Newton, an ESA science mission instrument, and contributions directly funded by ESA Member States and NASA. The authors thank Ileyk El Mellah for providing insightful comments. 
The research leading to these results has received funding from the European Union's Horizon 2020 Programme under the AHEAD2020 project (grant agreement n. 871158). 
EVL was partially supported by the Polish National Science Center under the grant No. 2021/41/B/ST9/04110.
BDM acknowledges support from the European Union's Horizon 2020 research and innovation programme under the Marie Sk{\l}odowska-Curie grant agreement No. 798726 and Ram\'on y Cajal Fellowship RYC2018-025950-I. AAZ and AR acknowledge support from the Polish National Science Center under the grants No. 2015/18/A/ST9/00746 and 2019/35/B/ST9/03944; and 2021/41/B/ST9/04110, respectively.  
TMB acknowledges financial contribution from the agreement ASI-INAF n.2017-14-H.0 and from PRIN-INAF 2019 N.15.
SM acknowledges financial support from the European Research Council (ERC) under the European Union's Horizon 2020 research and innovation program HotMilk (grant agreement No. 865637).

\section{DATA AVAILABILITY}
The data are publicly available from ESA's XMM-Newton Science Archive and NASA's HEASARC archive.



\clearpage 
\bibliographystyle{mnras}
\bibliography{bib} 

\begin{thebibliography}{}
\makeatletter
\relax
\def\mn@urlcharsother{\let\do\@makeother \do\$\do\&\do\#\do\^\do\_\do\%\do\~}
\def\mn@doi{\begingroup\mn@urlcharsother \@ifnextchar [ {\mn@doi@}
  {\mn@doi@[]}}
\def\mn@doi@[#1]#2{\def\@tempa{#1}\ifx\@tempa\@empty \href
  {http://dx.doi.org/#2} {doi:#2}\else \href {http://dx.doi.org/#2} {#1}\fi
  \endgroup}
\def\mn@eprint#1#2{\mn@eprint@#1:#2::\@nil}
\def\mn@eprint@arXiv#1{\href {http://arxiv.org/abs/#1} {{\tt arXiv:#1}}}
\def\mn@eprint@dblp#1{\href {http://dblp.uni-trier.de/rec/bibtex/#1.xml}
  {dblp:#1}}
\def\mn@eprint@#1:#2:#3:#4\@nil{\def\@tempa {#1}\def\@tempb {#2}\def\@tempc
  {#3}\ifx \@tempc \@empty \let \@tempc \@tempb \let \@tempb \@tempa \fi \ifx
  \@tempb \@empty \def\@tempb {arXiv}\fi \@ifundefined
  {mn@eprint@\@tempb}{\@tempb:\@tempc}{\expandafter \expandafter \csname
  mn@eprint@\@tempb\endcsname \expandafter{\@tempc}}}

\bibitem[\protect\citeauthoryear{{Ar{\'e}valo} \& {Uttley}}{{Ar{\'e}valo} \&
  {Uttley}}{2006}]{Arevalo_Uttley2006}
{Ar{\'e}valo} P.,  {Uttley} P.,  2006, \mn@doi [\mnras]
  {10.1111/j.1365-2966.2006.09989.x}, \href
  {https://ui.adsabs.harvard.edu/abs/2006MNRAS.367..801A} {367, 801}

\bibitem[\protect\citeauthoryear{{Arnaud}}{{Arnaud}}{1996}]{Arnaud_1996}
{Arnaud} K.~A.,  1996, in {Jacoby} G.~H.,  {Barnes} J.,  eds,  Astronomical
  Society of the Pacific Conference Series Vol. 101, Astronomical Data Analysis
  Software and Systems V. p.~17

\bibitem[\protect\citeauthoryear{{Axelsson} \& {Done}}{{Axelsson} \&
  {Done}}{2018}]{Axelsson_Done_2018}
{Axelsson} M.,  {Done} C.,  2018, \mn@doi [\mnras] {10.1093/mnras/sty1801},
  \href {https://ui.adsabs.harvard.edu/abs/2018MNRAS.480..751A} {480, 751}

\bibitem[\protect\citeauthoryear{{Axelsson}, {Borgonovo}  \&
  {Larsson}}{{Axelsson} et~al.}{2005}]{Axelsson_2005a}
{Axelsson} M.,  {Borgonovo} L.,   {Larsson} S.,  2005, \mn@doi [\aap]
  {10.1051/0004-6361:20042362}, \href
  {http://adsabs.harvard.edu/abs/2005A%26A...438..999A} {438, 999}

\bibitem[\protect\citeauthoryear{{Ba{\l}uci{\'n}ska-Church}, {Church},
  {Charles}, {Nagase}, {LaSala}  \& {Barnard}}{{Ba{\l}uci{\'n}ska-Church}
  et~al.}{2000}]{Balucinska_2000}
{Ba{\l}uci{\'n}ska-Church} M.,  {Church} M.~J.,  {Charles} P.~A.,  {Nagase} F.,
   {LaSala} J.,   {Barnard} R.,  2000, \mn@doi [\mnras]
  {10.1046/j.1365-8711.2000.03149.x}, \href
  {https://ui.adsabs.harvard.edu/abs/2000MNRAS.311..861B} {311, 861}

\bibitem[\protect\citeauthoryear{{Basak}, {Zdziarski}, {Parker}  \&
  {Islam}}{{Basak} et~al.}{2017}]{Basak_2017}
{Basak} R.,  {Zdziarski} A.~A.,  {Parker} M.,   {Islam} N.,  2017, \mn@doi
  [\mnras] {10.1093/mnras/stx2283}, \href
  {https://ui.adsabs.harvard.edu/abs/2017MNRAS.472.4220B} {472, 4220}

\bibitem[\protect\citeauthoryear{{Behar}, {Rasmussen}, {Blustin}, {Sako},
  {Kahn}, {Kaastra}, {Branduardi-Raymont}  \& {Steenbrugge}}{{Behar}
  et~al.}{2003}]{Behar_2003}
{Behar} E.,  {Rasmussen} A.~P.,  {Blustin} A.~J.,  {Sako} M.,  {Kahn} S.~M.,
  {Kaastra} J.~S.,  {Branduardi-Raymont} G.,   {Steenbrugge} K.~C.,  2003,
  \mn@doi [\apj] {10.1086/378853}, \href
  {https://ui.adsabs.harvard.edu/abs/2003ApJ...598..232B} {598, 232}

\bibitem[\protect\citeauthoryear{{Belloni}}{{Belloni}}{2010}]{Belloni_2010}
{Belloni} T.~M.,  2010, {States and Transitions in Black Hole Binaries}.
p.~53, \mn@doi{10.1007/978-3-540-76937-8\_3}

\bibitem[\protect\citeauthoryear{{Belloni} \& {Hasinger}}{{Belloni} \&
  {Hasinger}}{1990}]{Belloni_1990}
{Belloni} T.,  {Hasinger} G.,  1990, \aap, \href
  {https://ui.adsabs.harvard.edu/abs/1990A&A...227L..33B} {227, L33}

\bibitem[\protect\citeauthoryear{{Belloni}, {Hasinger}  \& {Kahabka}}{{Belloni}
  et~al.}{1991}]{Belloni_1991}
{Belloni} T.,  {Hasinger} G.,   {Kahabka} P.,  1991, \aap, \href
  {https://ui.adsabs.harvard.edu/abs/1991A&A...245L..29B} {245, L29}

\bibitem[\protect\citeauthoryear{{Belloni}, {Homan}, {Casella}, {van der Klis},
  {Nespoli}, {Lewin}, {Miller}  \& {M{\'e}ndez}}{{Belloni}
  et~al.}{2005}]{Belloni_2005}
{Belloni} T.,  {Homan} J.,  {Casella} P.,  {van der Klis} M.,  {Nespoli} E.,
  {Lewin} W.~H.~G.,  {Miller} J.~M.,   {M{\'e}ndez} M.,  2005, \mn@doi [\aap]
  {10.1051/0004-6361:20042457}, \href
  {https://ui.adsabs.harvard.edu/abs/2005A&A...440..207B} {440, 207}

\bibitem[\protect\citeauthoryear{{Belloni}, {Motta}  \&
  {Mu{\~n}oz-Darias}}{{Belloni} et~al.}{2011}]{Belloni_2011}
{Belloni} T.~M.,  {Motta} S.~E.,   {Mu{\~n}oz-Darias} T.,  2011, Bulletin of
  the Astronomical Society of India, \href
  {https://ui.adsabs.harvard.edu/abs/2011BASI...39..409B} {39, 409}

\bibitem[\protect\citeauthoryear{{Berger} \& {van der Klis}}{{Berger} \& {van
  der Klis}}{1994}]{Berger_vanderKliss_1994}
{Berger} M.,  {van der Klis} M.,  1994, \aap, \href
  {https://ui.adsabs.harvard.edu/abs/1994A&A...292..175B} {292, 175}

\bibitem[\protect\citeauthoryear{{B{\"o}ck} et~al.,}{{B{\"o}ck}
  et~al.}{2011}]{Boeck_2011a}
{B{\"o}ck} M.,  et~al., 2011, \mn@doi [\aap] {10.1051/0004-6361/201117159},
  \href {http://adsabs.harvard.edu/abs/2011A%26A...533A...8B} {533, A8+}

\bibitem[\protect\citeauthoryear{{Bollimpalli}, {Mahmoud}, {Done}, {Fragile},
  {Klu{\'z}niak}, {Narayan}  \& {White}}{{Bollimpalli}
  et~al.}{2020}]{Bollimpalli_2020}
{Bollimpalli} D.~A.,  {Mahmoud} R.,  {Done} C.,  {Fragile} P.~C.,
  {Klu{\'z}niak} W.,  {Narayan} R.,   {White} C.~J.,  2020, \mn@doi [\mnras]
  {10.1093/mnras/staa1808}, \href
  {https://ui.adsabs.harvard.edu/abs/2020MNRAS.496.3808B} {496, 3808}

\bibitem[\protect\citeauthoryear{{Boroson} \& {Vrtilek}}{{Boroson} \&
  {Vrtilek}}{2010}]{Boroson_Vrtilek_2010}
{Boroson} B.,  {Vrtilek} S.~D.,  2010, \mn@doi [\apj]
  {10.1088/0004-637X/710/1/197}, \href
  {https://ui.adsabs.harvard.edu/abs/2010ApJ...710..197B} {710, 197}

\bibitem[\protect\citeauthoryear{{Dauser}, {Garcia}, {Parker}, {Fabian}  \&
  {Wilms}}{{Dauser} et~al.}{2014}]{Dauser_2014}
{Dauser} T.,  {Garcia} J.,  {Parker} M.~L.,  {Fabian} A.~C.,   {Wilms} J.,
  2014, \mn@doi [\mnras] {10.1093/mnrasl/slu125}, \href
  {https://ui.adsabs.harvard.edu/abs/2014MNRAS.444L.100D} {444, L100}

\bibitem[\protect\citeauthoryear{{De Marco} \& {Ponti}}{{De Marco} \&
  {Ponti}}{2016}]{DeMarco_2016}
{De Marco} B.,  {Ponti} G.,  2016, \mn@doi [\apj] {10.3847/0004-637X/826/1/70},
  \href {https://ui.adsabs.harvard.edu/abs/2016ApJ...826...70D} {826, 70}

\bibitem[\protect\citeauthoryear{{De Marco}, {Ponti}, {Mu{\~n}oz-Darias}  \&
  {Nandra}}{{De Marco} et~al.}{2015}]{DeMarco_2015}
{De Marco} B.,  {Ponti} G.,  {Mu{\~n}oz-Darias} T.,   {Nandra} K.,  2015,
  \mn@doi [\apj] {10.1088/0004-637X/814/1/50}, \href
  {https://ui.adsabs.harvard.edu/abs/2015ApJ...814...50D} {814, 50}

\bibitem[\protect\citeauthoryear{{De Marco} et~al.,}{{De Marco}
  et~al.}{2017}]{DeMarco_2017}
{De Marco} B.,  et~al., 2017, \mn@doi [\mnras] {10.1093/mnras/stx1649}, \href
  {https://ui.adsabs.harvard.edu/abs/2017MNRAS.471.1475D} {471, 1475}

\bibitem[\protect\citeauthoryear{{De Marco} et~al.,}{{De Marco}
  et~al.}{2020}]{DeMarco_2020}
{De Marco} B.,  et~al., 2020, \mn@doi [\aap] {10.1051/0004-6361/201936470},
  \href {https://ui.adsabs.harvard.edu/abs/2020A&A...634A..65D} {634, A65}

\bibitem[\protect\citeauthoryear{{De Marco}, {Zdziarski}, {Ponti}, {Migliori},
  {Belloni}, {Segovia Otero}, {Dzie{\l}ak}  \& {Lai}}{{De Marco}
  et~al.}{2021}]{DeMarco_2021}
{De Marco} B.,  {Zdziarski} A.~A.,  {Ponti} G.,  {Migliori} G.,  {Belloni}
  T.~M.,  {Segovia Otero} A.,  {Dzie{\l}ak} M.~A.,   {Lai} E.~V.,  2021,
  \mn@doi [\aap] {10.1051/0004-6361/202140567}, \href
  {https://ui.adsabs.harvard.edu/abs/2021A&A...654A..14D} {654, A14}

\bibitem[\protect\citeauthoryear{{El Mellah}, {Grinberg}, {Sundqvist},
  {Driessen}  \& {Leutenegger}}{{El Mellah} et~al.}{2020}]{ElMellah_2020}
{El Mellah} I.,  {Grinberg} V.,  {Sundqvist} J.~O.,  {Driessen} F.~A.,
  {Leutenegger} M.~A.,  2020, \mn@doi [\aap] {10.1051/0004-6361/202038791},
  \href {https://ui.adsabs.harvard.edu/abs/2020A&A...643A...9E} {643, A9}

\bibitem[\protect\citeauthoryear{{Feldmeier}}{{Feldmeier}}{1995}]{Feldmeier_1995}
{Feldmeier} A.,  1995, \aap, \href
  {https://ui.adsabs.harvard.edu/abs/1995A&A...299..523F} {299, 523}

\bibitem[\protect\citeauthoryear{{Feng} \& {Cui}}{{Feng} \&
  {Cui}}{2002}]{Feng_Cui_2002}
{Feng} Y.~X.,  {Cui} W.,  2002, \mn@doi [\apj] {10.1086/324284}, \href
  {https://ui.adsabs.harvard.edu/abs/2002ApJ...564..953F} {564, 953}

\bibitem[\protect\citeauthoryear{{Garc{\'\i}a}, {Steiner}, {McClintock},
  {Remillard}, {Grinberg}  \& {Dauser}}{{Garc{\'\i}a}
  et~al.}{2015}]{Garcia_2015}
{Garc{\'\i}a} J.~A.,  {Steiner} J.~F.,  {McClintock} J.~E.,  {Remillard} R.~A.,
   {Grinberg} V.,   {Dauser} T.,  2015, \mn@doi [\apj]
  {10.1088/0004-637X/813/2/84}, \href
  {https://ui.adsabs.harvard.edu/abs/2015ApJ...813...84G} {813, 84}

\bibitem[\protect\citeauthoryear{{Gies} et~al.,}{{Gies}
  et~al.}{2003}]{Gies_2003}
{Gies} D.~R.,  et~al., 2003, \mn@doi [\apj] {10.1086/345345}, \href
  {https://ui.adsabs.harvard.edu/abs/2003ApJ...583..424G} {583, 424}

\bibitem[\protect\citeauthoryear{{Gies} et~al.,}{{Gies}
  et~al.}{2008}]{Gies_2008}
{Gies} D.~R.,  et~al., 2008, \mn@doi [\apj] {10.1086/586690}, \href
  {https://ui.adsabs.harvard.edu/abs/2008ApJ...678.1237G} {678, 1237}

\bibitem[\protect\citeauthoryear{{Gilfanov}}{{Gilfanov}}{2010}]{Gilfanov_2010}
{Gilfanov} M.,  2010, {X-Ray Emission from Black-Hole Binaries}.
p.~17, \mn@doi{10.1007/978-3-540-76937-8\_2}

\bibitem[\protect\citeauthoryear{{Grinberg} et~al.,}{{Grinberg}
  et~al.}{2014}]{Grinberg_2014}
{Grinberg} V.,  et~al., 2014, \mn@doi [\aap] {10.1051/0004-6361/201322969},
  \href {https://ui.adsabs.harvard.edu/abs/2014A&A...565A...1G} {565, A1}

\bibitem[\protect\citeauthoryear{{Grinberg} et~al.,}{{Grinberg}
  et~al.}{2015}]{Grinberg_2015}
{Grinberg} V.,  et~al., 2015, \mn@doi [\aap] {10.1051/0004-6361/201425418},
  \href {https://ui.adsabs.harvard.edu/abs/2015A&A...576A.117G} {576, A117}

\bibitem[\protect\citeauthoryear{{Grinberg} et~al.,}{{Grinberg}
  et~al.}{2017}]{Grinberg_2017}
{Grinberg} V.,  et~al., 2017, \mn@doi [\aap] {10.1051/0004-6361/201731843},
  \href {https://ui.adsabs.harvard.edu/abs/2017A&A...608A.143G} {608, A143}

\bibitem[\protect\citeauthoryear{{Grinberg}, {Nowak}  \& {Hell}}{{Grinberg}
  et~al.}{2020}]{Grinberg_2020}
{Grinberg} V.,  {Nowak} M.~A.,   {Hell} N.,  2020, \mn@doi [\aap]
  {10.1051/0004-6361/202039183}, \href
  {https://ui.adsabs.harvard.edu/abs/2020A&A...643A.109G} {643, A109}

\bibitem[\protect\citeauthoryear{{HI4PI Collaboration} et~al.,}{{HI4PI
  Collaboration} et~al.}{2016}]{HI4PIcoll}
{HI4PI Collaboration} et~al., 2016, \mn@doi [\aap]
  {10.1051/0004-6361/201629178}, \href
  {https://ui.adsabs.harvard.edu/abs/2016A&A...594A.116H} {594, A116}

\bibitem[\protect\citeauthoryear{{Hanke}, {Wilms}, {Nowak}, {Pottschmidt},
  {Schulz}  \& {Lee}}{{Hanke} et~al.}{2009}]{Hanke_2009a}
{Hanke} M.,  {Wilms} J.,  {Nowak} M.~A.,  {Pottschmidt} K.,  {Schulz} N.~S.,
  {Lee} J.~C.,  2009, \mn@doi [\apj] {10.1088/0004-637X/690/1/330}, \href
  {http://adsabs.harvard.edu/abs/2009ApJ...690..330H} {690, 330}

\bibitem[\protect\citeauthoryear{{Heil}, {Vaughan}  \& {Uttley}}{{Heil}
  et~al.}{2012}]{Heil_2012}
{Heil} L.~M.,  {Vaughan} S.,   {Uttley} P.,  2012, \mn@doi [\mnras]
  {10.1111/j.1365-2966.2012.20824.x}, \href
  {https://ui.adsabs.harvard.edu/abs/2012MNRAS.422.2620H} {422, 2620}

\bibitem[\protect\citeauthoryear{{Hirsch} et~al.,}{{Hirsch}
  et~al.}{2019}]{Hirsch_2019}
{Hirsch} M.,  et~al., 2019, \mn@doi [\aap] {10.1051/0004-6361/201935074}, \href
  {https://ui.adsabs.harvard.edu/abs/2019A&A...626A..64H} {626, A64}

\bibitem[\protect\citeauthoryear{{Homan} \& {Belloni}}{{Homan} \&
  {Belloni}}{2005}]{Homan_2005}
{Homan} J.,  {Belloni} T.,  2005, \mn@doi [\apss] {10.1007/s10509-005-1197-4},
  \href {https://ui.adsabs.harvard.edu/abs/2005Ap&SS.300..107H} {300, 107}

\bibitem[\protect\citeauthoryear{{Ibragimov}, {Poutanen}, {Gilfanov},
  {Zdziarski}  \& {Shrader}}{{Ibragimov} et~al.}{2005}]{Ibragimov_2005}
{Ibragimov} A.,  {Poutanen} J.,  {Gilfanov} M.,  {Zdziarski} A.~A.,   {Shrader}
  C.~R.,  2005, \mn@doi [\mnras] {10.1111/j.1365-2966.2005.09415.x}, \href
  {https://ui.adsabs.harvard.edu/abs/2005MNRAS.362.1435I} {362, 1435}

\bibitem[\protect\citeauthoryear{{Ichimaru}}{{Ichimaru}}{1977}]{Ichimaru_1977}
{Ichimaru} S.,  1977, \mn@doi [\apj] {10.1086/155314}, \href
  {https://ui.adsabs.harvard.edu/abs/1977ApJ...214..840I} {214, 840}

\bibitem[\protect\citeauthoryear{{Ingram}}{{Ingram}}{2019}]{Ingram_2019}
{Ingram} A.,  2019, \mn@doi [\mnras] {10.1093/mnras/stz2409}, \href
  {https://ui.adsabs.harvard.edu/abs/2019MNRAS.489.3927I} {489, 3927}

\bibitem[\protect\citeauthoryear{{Ingram} \& {van der Klis}}{{Ingram} \& {van
  der Klis}}{2013}]{Ingram_vanderKlis2013}
{Ingram} A.,  {van der Klis} M.,  2013, \mn@doi [\mnras]
  {10.1093/mnras/stt1107}, \href
  {https://ui.adsabs.harvard.edu/abs/2013MNRAS.434.1476I} {434, 1476}

\bibitem[\protect\citeauthoryear{{Jur{\'a}{\v{n}}ov{\'a}}, {Costantini}  \&
  {Uttley}}{{Jur{\'a}{\v{n}}ov{\'a}} et~al.}{2022}]{Juranova_2022}
{Jur{\'a}{\v{n}}ov{\'a}} A.,  {Costantini} E.,   {Uttley} P.,  2022, \mn@doi
  [\mnras] {10.1093/mnras/stab3731}, \href
  {https://ui.adsabs.harvard.edu/abs/2022MNRAS.510.4225J} {510, 4225}

\bibitem[\protect\citeauthoryear{{Kaastra} et~al.,}{{Kaastra}
  et~al.}{2012}]{Kaastra_2012}
{Kaastra} J.~S.,  et~al., 2012, \mn@doi [\aap] {10.1051/0004-6361/201118161},
  \href {https://ui.adsabs.harvard.edu/abs/2012A&A...539A.117K} {539, A117}

\bibitem[\protect\citeauthoryear{{Kara} et~al.,}{{Kara}
  et~al.}{2019}]{Kara_2019}
{Kara} E.,  et~al., 2019, \mn@doi [\nat] {10.1038/s41586-018-0803-x}, \href
  {https://ui.adsabs.harvard.edu/abs/2019Natur.565..198K} {565, 198}

\bibitem[\protect\citeauthoryear{{Kotov}, {Churazov}  \& {Gilfanov}}{{Kotov}
  et~al.}{2001}]{Kotov_2001}
{Kotov} O.,  {Churazov} E.,   {Gilfanov} M.,  2001, \mn@doi [\mnras]
  {10.1046/j.1365-8711.2001.04769.x}, \href
  {https://ui.adsabs.harvard.edu/abs/2001MNRAS.327..799K} {327, 799}

\bibitem[\protect\citeauthoryear{{Krongold}, {Nicastro}, {Elvis}, {Brickhouse},
  {Binette}, {Mathur}  \& {Jim{\'e}nez-Bail{\'o}n}}{{Krongold}
  et~al.}{2007}]{Krongold_2007}
{Krongold} Y.,  {Nicastro} F.,  {Elvis} M.,  {Brickhouse} N.,  {Binette} L.,
  {Mathur} S.,   {Jim{\'e}nez-Bail{\'o}n} E.,  2007, \mn@doi [\apj]
  {10.1086/512476}, \href
  {https://ui.adsabs.harvard.edu/abs/2007ApJ...659.1022K} {659, 1022}

\bibitem[\protect\citeauthoryear{{Li} \& {Clark}}{{Li} \&
  {Clark}}{1974}]{Li_1974}
{Li} F.~K.,  {Clark} G.~W.,  1974, \mn@doi [\apjl] {10.1086/181538}, \href
  {https://ui.adsabs.harvard.edu/abs/1974ApJ...191L..27L} {191, L27}

\bibitem[\protect\citeauthoryear{{Ling}, {Zhang}, {Xiang}  \& {Tang}}{{Ling}
  et~al.}{2009}]{Ling_2009}
{Ling} Z.,  {Zhang} S.~N.,  {Xiang} J.,   {Tang} S.,  2009, \mn@doi [\apj]
  {10.1088/0004-637X/690/1/224}, \href
  {https://ui.adsabs.harvard.edu/abs/2009ApJ...690..224L} {690, 224}

\bibitem[\protect\citeauthoryear{{Lyubarskii}}{{Lyubarskii}}{1997}]{Lyubarskii_1997}
{Lyubarskii} Y.~E.,  1997, \mn@doi [\mnras] {10.1093/mnras/285.3.604}, \href
  {https://ui.adsabs.harvard.edu/abs/1997MNRAS.285..604L} {285, 604}

\bibitem[\protect\citeauthoryear{{Mahmoud} \& {Done}}{{Mahmoud} \&
  {Done}}{2018a}]{Mahmoud_2018_A}
{Mahmoud} R.~D.,  {Done} C.,  2018a, \mn@doi [\mnras] {10.1093/mnras/stx2359},
  \href {https://ui.adsabs.harvard.edu/abs/2018MNRAS.473.2084M} {473, 2084}

\bibitem[\protect\citeauthoryear{{Mahmoud} \& {Done}}{{Mahmoud} \&
  {Done}}{2018b}]{Mahmoud_2018_B}
{Mahmoud} R.~D.,  {Done} C.,  2018b, \mn@doi [\mnras] {10.1093/mnras/sty2133},
  \href {https://ui.adsabs.harvard.edu/abs/2018MNRAS.480.4040M} {480, 4040}

\bibitem[\protect\citeauthoryear{{Miller-Jones} et~al.,}{{Miller-Jones}
  et~al.}{2021}]{MillerJones_2021}
{Miller-Jones} J. C.~A.,  et~al., 2021, \mn@doi [Science]
  {10.1126/science.abb3363}, \href
  {https://ui.adsabs.harvard.edu/abs/2021Sci...371.1046M} {371, 1046}

\bibitem[\protect\citeauthoryear{{Miller}, {Wojdowski}, {Schulz}, {Marshall},
  {Fabian}, {Remillard}, {Wijnands}  \& {Lewin}}{{Miller}
  et~al.}{2005}]{Miller_2005a}
{Miller} J.~M.,  {Wojdowski} P.,  {Schulz} N.~S.,  {Marshall} H.~L.,  {Fabian}
  A.~C.,  {Remillard} R.~A.,  {Wijnands} R.,   {Lewin} W.~H.~G.,  2005, \mn@doi
  [\apj] {10.1086/426701}, \href
  {http://adsabs.harvard.edu/abs/2005ApJ...620..398M} {620, 398}

\bibitem[\protect\citeauthoryear{{Misra}}{{Misra}}{2000}]{Misra_2000}
{Misra} R.,  2000, \mn@doi [\apjl] {10.1086/312470}, \href
  {https://ui.adsabs.harvard.edu/abs/2000ApJ...529L..95M} {529, L95}

\bibitem[\protect\citeauthoryear{{Mitsuda} et~al.,}{{Mitsuda}
  et~al.}{1984}]{Mitsuda_1984}
{Mitsuda} K.,  et~al., 1984, \pasj, \href
  {https://ui.adsabs.harvard.edu/abs/1984PASJ...36..741M} {36, 741}

\bibitem[\protect\citeauthoryear{{Mi{\v{s}}kovi{\v{c}}ov{\'a}}
  et~al.,}{{Mi{\v{s}}kovi{\v{c}}ov{\'a}} et~al.}{2016}]{Miskovicova_2016}
{Mi{\v{s}}kovi{\v{c}}ov{\'a}} I.,  et~al., 2016, \mn@doi [\aap]
  {10.1051/0004-6361/201322490}, \href
  {https://ui.adsabs.harvard.edu/abs/2016A&A...590A.114M} {590, A114}

\bibitem[\protect\citeauthoryear{{Miyamoto}, {Kimura}, {Kitamoto}, {Dotani}  \&
  {Ebisawa}}{{Miyamoto} et~al.}{1991}]{Miyamoto_1991}
{Miyamoto} S.,  {Kimura} K.,  {Kitamoto} S.,  {Dotani} T.,   {Ebisawa} K.,
  1991, \mn@doi [\apj] {10.1086/170837}, \href
  {https://ui.adsabs.harvard.edu/abs/1991ApJ...383..784M} {383, 784}

\bibitem[\protect\citeauthoryear{{Mu{\~n}oz-Darias}, {Motta}  \&
  {Belloni}}{{Mu{\~n}oz-Darias} et~al.}{2011}]{MunozDarias_2011}
{Mu{\~n}oz-Darias} T.,  {Motta} S.,   {Belloni} T.~M.,  2011, \mn@doi [\mnras]
  {10.1111/j.1365-2966.2010.17476.x}, \href
  {https://ui.adsabs.harvard.edu/abs/2011MNRAS.410..679M} {410, 679}

\bibitem[\protect\citeauthoryear{{Mushtukov}, {Ingram}  \& {van der
  Klis}}{{Mushtukov} et~al.}{2018}]{Mushtukov_2018}
{Mushtukov} A.~A.,  {Ingram} A.,   {van der Klis} M.,  2018, \mn@doi [\mnras]
  {10.1093/mnras/stx2872}, \href
  {https://ui.adsabs.harvard.edu/abs/2018MNRAS.474.2259M} {474, 2259}

\bibitem[\protect\citeauthoryear{{Narayan} \& {Yi}}{{Narayan} \&
  {Yi}}{1994}]{Narayan_Yi_1994}
{Narayan} R.,  {Yi} I.,  1994, \mn@doi [\apjl] {10.1086/187381}, \href
  {https://ui.adsabs.harvard.edu/abs/1994ApJ...428L..13N} {428, L13}

\bibitem[\protect\citeauthoryear{{Ng}, {D{\'\i}az Trigo}, {Cadolle Bel}  \&
  {Migliari}}{{Ng} et~al.}{2010}]{Ng_2010}
{Ng} C.,  {D{\'\i}az Trigo} M.,  {Cadolle Bel} M.,   {Migliari} S.,  2010,
  \mn@doi [\aap] {10.1051/0004-6361/200913575}, \href
  {https://ui.adsabs.harvard.edu/abs/2010A&A...522A..96N} {522, A96}

\bibitem[\protect\citeauthoryear{{Nicastro}, {Fiore}, {Perola}  \&
  {Elvis}}{{Nicastro} et~al.}{1999}]{Nicastro_1999}
{Nicastro} F.,  {Fiore} F.,  {Perola} G.~C.,   {Elvis} M.,  1999, \mn@doi
  [\apj] {10.1086/306736}, \href
  {https://ui.adsabs.harvard.edu/abs/1999ApJ...512..184N} {512, 184}

\bibitem[\protect\citeauthoryear{{Nowak}, {Vaughan}, {Wilms}, {Dove}  \&
  {Begelman}}{{Nowak} et~al.}{1999}]{Nowak_1999}
{Nowak} M.~A.,  {Vaughan} B.~A.,  {Wilms} J.,  {Dove} J.~B.,   {Begelman}
  M.~C.,  1999, \mn@doi [\apj] {10.1086/306610}, \href
  {https://ui.adsabs.harvard.edu/abs/1999ApJ...510..874N} {510, 874}

\bibitem[\protect\citeauthoryear{{Nowak} et~al.,}{{Nowak}
  et~al.}{2011}]{Nowak_2011}
{Nowak} M.~A.,  et~al., 2011, \mn@doi [\apj] {10.1088/0004-637X/728/1/13},
  \href {https://ui.adsabs.harvard.edu/abs/2011ApJ...728...13N} {728, 13}

\bibitem[\protect\citeauthoryear{{Owocki} \& {Rybicki}}{{Owocki} \&
  {Rybicki}}{1984}]{Owocki_1984}
{Owocki} S.~P.,  {Rybicki} G.~B.,  1984, \mn@doi [\apj] {10.1086/162412}, \href
  {https://ui.adsabs.harvard.edu/abs/1984ApJ...284..337O} {284, 337}

\bibitem[\protect\citeauthoryear{{Owocki}, {Castor}  \& {Rybicki}}{{Owocki}
  et~al.}{1988}]{Owocki_1988}
{Owocki} S.~P.,  {Castor} J.~I.,   {Rybicki} G.~B.,  1988, \mn@doi [\apj]
  {10.1086/166977}, \href
  {https://ui.adsabs.harvard.edu/abs/1988ApJ...335..914O} {335, 914}

\bibitem[\protect\citeauthoryear{{Pottschmidt}, {Wilms}, {Nowak}, {Heindl},
  {Smith}  \& {Staubert}}{{Pottschmidt} et~al.}{2000}]{Pottschimdt_2000}
{Pottschmidt} K.,  {Wilms} J.,  {Nowak} M.~A.,  {Heindl} W.~A.,  {Smith} D.~M.,
    {Staubert} R.,  2000, \aap, \href
  {https://ui.adsabs.harvard.edu/abs/2000A&A...357L..17P} {357, L17}

\bibitem[\protect\citeauthoryear{{Pottschmidt} et~al.,}{{Pottschmidt}
  et~al.}{2003}]{Pottschmidt_2003}
{Pottschmidt} K.,  et~al., 2003, \mn@doi [\aap] {10.1051/0004-6361:20030906},
  \href {https://ui.adsabs.harvard.edu/abs/2003A&A...407.1039P} {407, 1039}

\bibitem[\protect\citeauthoryear{{Poutanen} \& {Fabian}}{{Poutanen} \&
  {Fabian}}{1999}]{Poutanen_1999}
{Poutanen} J.,  {Fabian} A.~C.,  1999, \mn@doi [\mnras]
  {10.1046/j.1365-8711.1999.02735.x}, \href
  {https://ui.adsabs.harvard.edu/abs/1999MNRAS.306L..31P} {306, L31}

\bibitem[\protect\citeauthoryear{{Poutanen}, {Zdziarski}  \&
  {Ibragimov}}{{Poutanen} et~al.}{2008}]{Poutanen_2008}
{Poutanen} J.,  {Zdziarski} A.~A.,   {Ibragimov} A.,  2008, \mn@doi [\mnras]
  {10.1111/j.1365-2966.2008.13666.x}, \href
  {https://ui.adsabs.harvard.edu/abs/2008MNRAS.389.1427P} {389, 1427}

\bibitem[\protect\citeauthoryear{{Predehl} \& {Schmitt}}{{Predehl} \&
  {Schmitt}}{1995}]{Predehl_Schmidt_1995}
{Predehl} P.,  {Schmitt} J.~H.~M.~M.,  1995, \aap, \href
  {https://ui.adsabs.harvard.edu/abs/1995A&A...293..889P} {500, 459}

\bibitem[\protect\citeauthoryear{{Remillard} \& {Canizares}}{{Remillard} \&
  {Canizares}}{1984}]{Remillard_1984}
{Remillard} R.~A.,  {Canizares} C.~R.,  1984, \mn@doi [\apj] {10.1086/161846},
  \href {https://ui.adsabs.harvard.edu/abs/1984ApJ...278..761R} {278, 761}

\bibitem[\protect\citeauthoryear{{Shakura} \& {Sunyaev}}{{Shakura} \&
  {Sunyaev}}{1973}]{Shakura_Sunyaev_1973}
{Shakura} N.~I.,  {Sunyaev} R.~A.,  1973, \aap, \href
  {https://ui.adsabs.harvard.edu/abs/1973A&A....24..337S} {500, 33}

\bibitem[\protect\citeauthoryear{{Shapiro}, {Lightman}  \& {Eardley}}{{Shapiro}
  et~al.}{1976}]{Shapiro_1976}
{Shapiro} S.~L.,  {Lightman} A.~P.,   {Eardley} D.~M.,  1976, \mn@doi [\apj]
  {10.1086/154162}, \href
  {https://ui.adsabs.harvard.edu/abs/1976ApJ...204..187S} {204, 187}

\bibitem[\protect\citeauthoryear{{Silva}, {Uttley}  \& {Costantini}}{{Silva}
  et~al.}{2016}]{Silva_2016}
{Silva} C.~V.,  {Uttley} P.,   {Costantini} E.,  2016, \mn@doi [\aap]
  {10.1051/0004-6361/201628555}, \href
  {https://ui.adsabs.harvard.edu/abs/2016A&A...596A..79S} {596, A79}

\bibitem[\protect\citeauthoryear{{Sunyaev} \& {Truemper}}{{Sunyaev} \&
  {Truemper}}{1979}]{Sunyaev_1979}
{Sunyaev} R.~A.,  {Truemper} J.,  1979, \mn@doi [\nat] {10.1038/279506a0},
  \href {https://ui.adsabs.harvard.edu/abs/1979Natur.279..506S} {279, 506}

\bibitem[\protect\citeauthoryear{{Uttley}, {Wilkinson}, {Cassatella}, {Wilms},
  {Pottschmidt}, {Hanke}  \& {B{\"o}ck}}{{Uttley} et~al.}{2011}]{Uttley_2011}
{Uttley} P.,  {Wilkinson} T.,  {Cassatella} P.,  {Wilms} J.,  {Pottschmidt} K.,
   {Hanke} M.,   {B{\"o}ck} M.,  2011, \mn@doi [\mnras]
  {10.1111/j.1745-3933.2011.01056.x}, \href
  {https://ui.adsabs.harvard.edu/abs/2011MNRAS.414L..60U} {414, L60}

\bibitem[\protect\citeauthoryear{{Uttley}, {Cackett}, {Fabian}, {Kara}  \&
  {Wilkins}}{{Uttley} et~al.}{2014}]{Uttley_2014}
{Uttley} P.,  {Cackett} E.~M.,  {Fabian} A.~C.,  {Kara} E.,   {Wilkins} D.~R.,
  2014, \mn@doi [\aapr] {10.1007/s00159-014-0072-0}, \href
  {https://ui.adsabs.harvard.edu/abs/2014A&ARv..22...72U} {22, 72}

\bibitem[\protect\citeauthoryear{{Vaughan} \& {Nowak}}{{Vaughan} \&
  {Nowak}}{1997}]{Vaughan_Nowak1997}
{Vaughan} B.~A.,  {Nowak} M.~A.,  1997, \mn@doi [\apjl] {10.1086/310430}, \href
  {https://ui.adsabs.harvard.edu/abs/Vaughan_Nowak1997} {474, L43}

\bibitem[\protect\citeauthoryear{{Veledina}, {Poutanen}  \& {Vurm}}{{Veledina}
  et~al.}{2011}]{Veledina_2011}
{Veledina} A.,  {Poutanen} J.,   {Vurm} I.,  2011, \mn@doi [\apjl]
  {10.1088/2041-8205/737/1/L17}, \href
  {https://ui.adsabs.harvard.edu/abs/2011ApJ...737L..17V} {737, L17}

\bibitem[\protect\citeauthoryear{{Vurm} \& {Poutanen}}{{Vurm} \&
  {Poutanen}}{2008}]{Vurm_Poutanen2008}
{Vurm} I.,  {Poutanen} J.,  2008, \mn@doi [International Journal of Modern
  Physics D] {10.1142/S0218271808013236}, \href
  {https://ui.adsabs.harvard.edu/abs/2008IJMPD..17.1629V} {17, 1629}

\bibitem[\protect\citeauthoryear{{Wang} et~al.,}{{Wang}
  et~al.}{2020}]{Wang_2020}
{Wang} J.,  et~al., 2020, \mn@doi [\apj] {10.3847/1538-4357/ab9ec3}, \href
  {https://ui.adsabs.harvard.edu/abs/2020ApJ...899...44W} {899, 44}

\bibitem[\protect\citeauthoryear{{Wilkinson} \& {Uttley}}{{Wilkinson} \&
  {Uttley}}{2009}]{Wilkinson_Uttley2009}
{Wilkinson} T.,  {Uttley} P.,  2009, \mn@doi [\mnras]
  {10.1111/j.1365-2966.2009.15008.x}, \href
  {https://ui.adsabs.harvard.edu/abs/2009MNRAS.397..666W} {397, 666}

\bibitem[\protect\citeauthoryear{{Wilms}, {Allen}  \& {McCray}}{{Wilms}
  et~al.}{2000}]{Wilms_2000}
{Wilms} J.,  {Allen} A.,   {McCray} R.,  2000, \mn@doi [\apj] {10.1086/317016},
  \href {https://ui.adsabs.harvard.edu/abs/2000ApJ...542..914W} {542, 914}

\bibitem[\protect\citeauthoryear{{Wilms}, {Nowak}, {Pottschmidt}, {Pooley}  \&
  {Fritz}}{{Wilms} et~al.}{2006}]{Wilms_2006}
{Wilms} J.,  {Nowak} M.~A.,  {Pottschmidt} K.,  {Pooley} G.~G.,   {Fritz} S.,
  2006, \mn@doi [\aap] {10.1051/0004-6361:20053938}, \href
  {https://ui.adsabs.harvard.edu/abs/2006A&A...447..245W} {447, 245}

\bibitem[\protect\citeauthoryear{{Xiang}, {Lee}, {Nowak}  \& {Wilms}}{{Xiang}
  et~al.}{2011}]{Xiang_2011}
{Xiang} J.,  {Lee} J.~C.,  {Nowak} M.~A.,   {Wilms} J.,  2011, \mn@doi [\apj]
  {10.1088/0004-637X/738/1/78}, \href
  {https://ui.adsabs.harvard.edu/abs/2011ApJ...738...78X} {738, 78}

\bibitem[\protect\citeauthoryear{{Zdziarski}}{{Zdziarski}}{2012}]{Zdziarski_2012}
{Zdziarski} A.~A.,  2012, \mn@doi [\mnras] {10.1111/j.1365-2966.2012.20754.x},
  \href {https://ui.adsabs.harvard.edu/abs/2012MNRAS.422.1750Z} {422, 1750}

\bibitem[\protect\citeauthoryear{{Zdziarski} \& {Gierli{\'n}ski}}{{Zdziarski}
  \& {Gierli{\'n}ski}}{2004}]{Zdziarski_2004}
{Zdziarski} A.~A.,  {Gierli{\'n}ski} M.,  2004, \mn@doi [Progress of
  Theoretical Physics Supplement] {10.1143/PTPS.155.99}, \href
  {https://ui.adsabs.harvard.edu/abs/2004PThPS.155...99Z} {155, 99}

\bibitem[\protect\citeauthoryear{{Zdziarski}, {Johnson}  \&
  {Magdziarz}}{{Zdziarski} et~al.}{1996}]{Zdziarski_1996}
{Zdziarski} A.~A.,  {Johnson} W.~N.,   {Magdziarz} P.,  1996, \mn@doi [\mnras]
  {10.1093/mnras/283.1.193}, \href
  {https://ui.adsabs.harvard.edu/abs/1996MNRAS.283..193Z} {283, 193}

\bibitem[\protect\citeauthoryear{{Zdziarski}, {Misra}  \&
  {Gierli{\'n}ski}}{{Zdziarski} et~al.}{2010}]{Zdziarski_2010}
{Zdziarski} A.~A.,  {Misra} R.,   {Gierli{\'n}ski} M.,  2010, \mn@doi [\mnras]
  {10.1111/j.1365-2966.2009.15942.x}, \href
  {https://ui.adsabs.harvard.edu/abs/2010MNRAS.402..767Z} {402, 767}

\bibitem[\protect\citeauthoryear{{{\.Z}ycki}, {Done}  \& {Smith}}{{{\.Z}ycki}
  et~al.}{1999}]{Zycki_1999}
{{\.Z}ycki} P.~T.,  {Done} C.,   {Smith} D.~A.,  1999, \mn@doi [\mnras]
  {10.1046/j.1365-8711.1999.02885.x}, \href
  {https://ui.adsabs.harvard.edu/abs/1999MNRAS.309..561Z} {309, 561}

\makeatother
\end{thebibliography}







\clearpage 
\appendix
\section{Modeling the colour-colour diagram of observation 201}\label{APP_CCDiagram_model_data}
We modeled the colour-colour diagram of each observation of Cyg X-1 with a simple absorption model. As an example, here we show the results obtained for observation 201.
We used Xspec to model the continuum using a partially absorbed power law with spectral index 1.7 (\texttt{tbpcf} $\times$ \texttt{powerlaw}). We fixed the covering fraction at $0.9$ and let the wind column density $\mathrm{N_{H,w}}$ vary (from $0.1 \times 10^{22}\,\mathrm{cm^{-2}}$ to $3.5 \times 10^{23}\,\mathrm{cm^{-2}}$). This allowed us to build the colour-colour diagram track show in Fig.~\ref{fig:CCDiagram_appendix} and overplotted to the data. The model also includes ISM absorption (\texttt{TBnew}) with Galactic $\mathrm{N_H} = 0.7 \times 10^{22}\,\mathrm{cm^{-2}}$ \citep{Basak_2017,HI4PIcoll} where abundances from \cite{Wilms_2000} were used. 

For such a model, the threshold of hard and soft colours used in Sect.~\ref{selezione_NWA} to filter out data characterised by strong wind absorption, corresponds, for a covering factor of 0.9, to a wind column density of $\mathrm{N_{H,w}} \geqslant 1.08 \times 10^{22} \mathrm{cm^{-2}}$. 

\includegraphics[trim={2.3cm 0cm 0cm 0cm},width=0.48\textwidth]{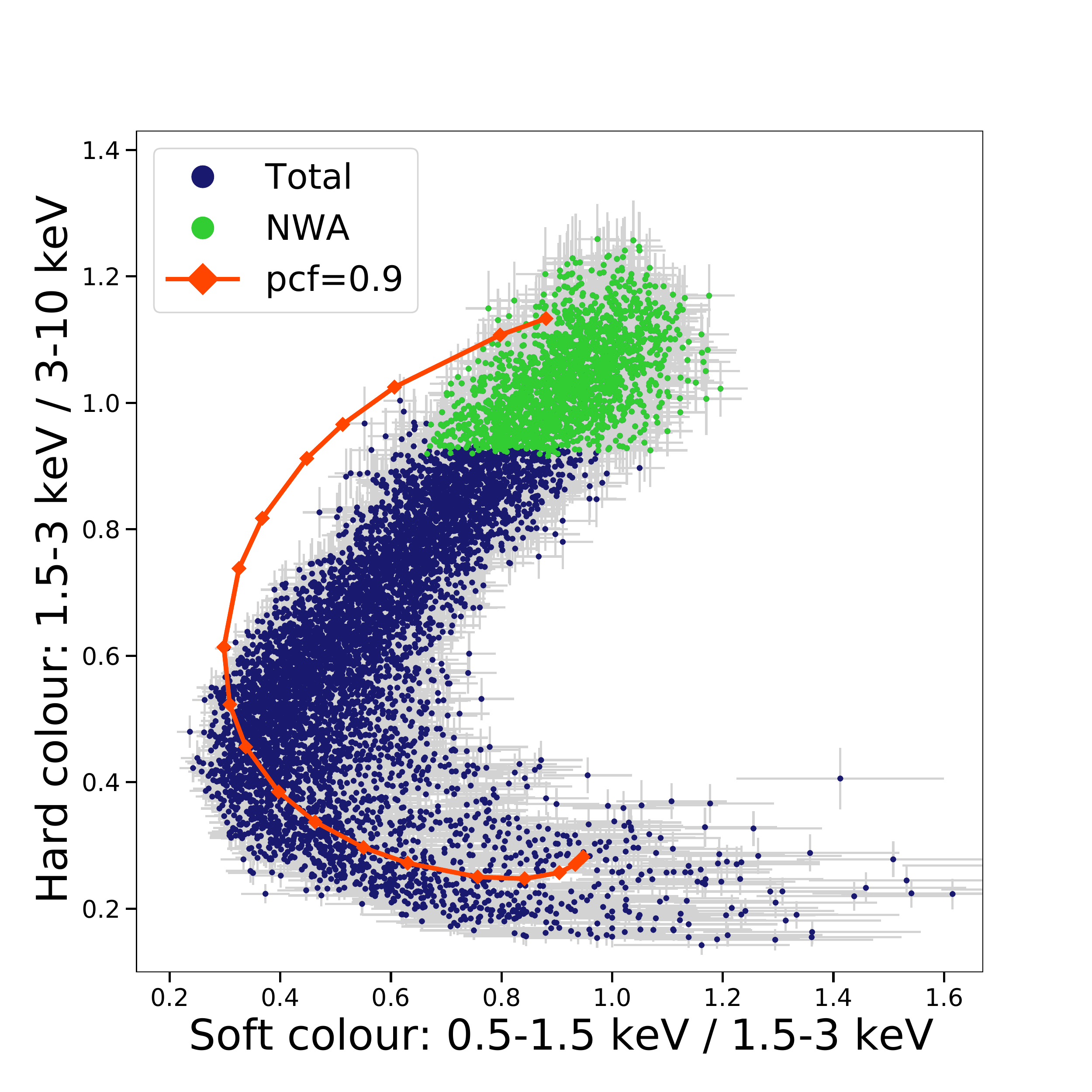}
\captionof{figure}{Colour-colour diagram of observation 201 with the simulated model (\texttt{TBnew} $\times$ \texttt{tbpcf} $\times$ \texttt{powerlaw} in Xspec) for a partial covering factor of 0.9 (red line), overplotted on the data. Each red point on the model curve corresponds to a different value of $\mathrm{N_{H, w}}$ (from $0.1 \times 10^{22}\,\mathrm{cm^{-2}}$, first point on the top, to $3.5 \times 10^{23}\,\mathrm{cm^{-2}}$, last point on the bottom, arbitrarily spaced).
The green points correspond to the selected NWA dataset.}
\label{fig:CCDiagram_appendix}

\section{Selection of the NWA dataset for observation 701}\label{APP_701}
\begin{figure*}
\includegraphics[trim={4cm 0 8cm 0},width=1\linewidth]{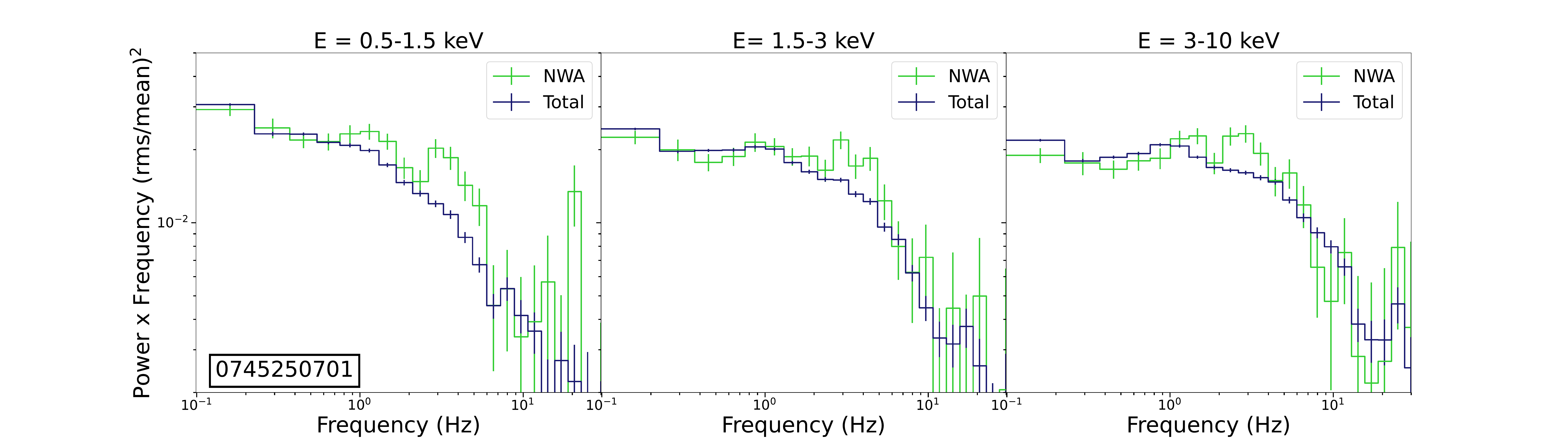}
\captionof{figure}{PSD for Total (blue) and NWA (green) light curves in the soft (0.5--1.5~keV), intermediate (1.5--3~keV) and hard (3--10~keV) band for observation 701 after selection of datasets characterised by hard colour $\geqslant$~1.05 and soft colour $\geqslant$~1.}
\label{fig:APP_701}
\end{figure*}

We verified that a stricter (from that used in Sect.~\ref{selezione_NWA}) selection of hard and soft colours allows recovering the double-hump shape of the PSD of observation 701 (corresponding to the second passage at superior conjunction). For this we chose hard colours $\geqslant$~1.05 and soft colours $\geqslant$~1.  
Results are shown in Fig.~\ref{fig:APP_701}. 

\section{Spectral fitting parameters}\label{tab_APP}
We report (Table \ref{tab_spectra}) the broad band continuum best-fit parameters for the model \texttt{Tbnew $\times$ [diskbb + nthComp + relxillCp]} fitted jointly to the spectrum of each XMM-Newton observation of Cyg X-1 (as described in Sect.~\ref{CS}).

\setlength{\tabcolsep}{18pt}
\renewcommand{\arraystretch}{1.2}
\begin{table*}
    \caption{Best-fit parameters for each XMM-Newton observation of Cyg X-1. The $kT_{in}$ and $kT_{e}$ parameters are, respectively, the inner temperature of the disc and the electron temperature of the soft Comptonisation component. The $\Gamma_{H}$ and $\Gamma_{S}$ parameters are, respectively, the spectral index of the hard and soft Comptonisation components.
    Errors are indicated with a confidence level of 90\%. Unconstrained parameters (indicated with $f$) were kept fixed at the best-fit value.}
	\label{tab_spectra}
	\begin{tabular}{ccccccc}
		\hline
		Component & Parameter & 201 & 501 & $601_\mathrm{A}$ &$601_\mathrm{B}$ & 701 \\
		   
		\hline
		
		\texttt{diskbb}  & $kT_{in}$ (keV) & $0.18_{-0.01}^{+0.01}$
		                             & $0.19_{-0.01}^{+0.01}$
		                             & $0.17_{-0.01}^{+0.01}$
		                             & $0.17_{-0.01}^{+0.01}$
		                             & $0.20_{-0.01}^{+0.01}$\\
        	
		\hline
		\texttt{nthComp}     & $\Gamma_{S}$ & $2.02_{-0.04}^{+0.04}$
		                                         & $1.68_{-0.02}^{+0.08}$
		                                         & $2.23_{-0.03}^{+0.01}$
		                                         & $2.51_{-0.02}^{+0.09}$
		                                         & $1.35_{-0.04}^{+0.08}$\\
		                                        
		                    & $kT_{e}$ (keV) & $0.80_{-0.04}^{+0.02}$ 
		                            & $1^f$
		                            & $1^f$
		                            & $1^f$
		                            & $<0.77$\\
        
        \hline
		\texttt{relxillCp} & $\Gamma_{H}$ & $1.38_{-0.02}^{+0.02}$ 
		                             & $1.32_{-0.02}^{+0.04}$ 
		                             & $1.35_{-0.02}^{+0.02}$
		                             & $1.42_{-0.05}^{+0.04}$  
		                             & $1.38_{-0.01}^{+0.02}$ \\
	    \hline
	\end{tabular} 

\end{table*}

\bsp	
\label{lastpage}
\end{document}